\documentclass[10pt,twocolumn,letterpaper]{article}

\usepackage{titling}
\usepackage{iccv}
\usepackage{times}
\usepackage{epsfig}
\usepackage{graphicx}
\usepackage{amsmath}
\usepackage{amssymb}
\usepackage{booktabs}
\usepackage{subfigure}
\usepackage{multirow}
\usepackage[accsupp]{axessibility}

\usepackage[breaklinks=true,bookmarks=false]{hyperref}

\iccvfinalcopy 

\def\iccvPaperID{3772} 

\ificcvfinal\fi

\newcommand{\beginsupplement}{%
        \setcounter{table}{0}
        \renewcommand{\thetable}{A\arabic{table}}%
        \setcounter{figure}{0}
        \renewcommand{\thefigure}{A\arabic{figure}}%
        \setcounter{section}{0}
        \renewcommand{\thesection}{A\arabic{section}}%
}

\usepackage{overpic}
\usepackage{enumitem} 
\usepackage{overpic} 
\usepackage{color}

\definecolor{turquoise}{cmyk}{0.65,0,0.1,0.3}
\definecolor{purple}{rgb}{0.65,0,0.65}
\definecolor{dark_green}{rgb}{0, 0.5, 0}
\definecolor{orange}{rgb}{0.8, 0.6, 0.2}
\definecolor{red}{rgb}{0.8, 0.2, 0.2}
\definecolor{darkred}{rgb}{0.6, 0.1, 0.05}
\definecolor{blueish}{rgb}{0.0, 0.3, .6}
\definecolor{light_gray}{rgb}{0.7, 0.7, .7}
\definecolor{pink}{rgb}{1, 0, 1}
\definecolor{greyblue}{rgb}{0.25, 0.25, 1}

\definecolor{light-gray}{gray}{0.70}

\begin{document}

\predate{}
\postdate{}

\title{\textbf{TransTIC: Transferring Transformer-based Image Compression \\ from Human Perception to Machine Perception}}

\author{
Yi-Hsin Chen \quad Ying-Chieh Weng \quad Chia-Hao Kao \quad Cheng Chien \\ Wei-Chen Chiu \quad Wen-Hsiao Peng \\
National Yang Ming Chiao Tung University, Taiwan\\
\tt\small \{yhchen12101.cs09@, wengyc.cs09@, chiahaok.cs10@, cchien1999@cs.\}nycu.edu.tw \\ \tt\small \{walon, wpeng\}@cs.nctu.edu.tw
}

\date{}

\makeatletter
\def\thanks#1{\protected@xdef\@thanks{\@thanks
        \protect\footnotetext{#1}}}
\makeatother

\maketitle
\ificcvfinal\thispagestyle{empty}\fi

\begin{abstract}
This work aims for transferring a Transformer-based image compression codec from human perception to machine perception without fine-tuning the codec. We propose a transferable Transformer-based image compression framework, termed TransTIC. Inspired by visual prompt tuning, TransTIC adopts an instance-specific prompt generator to inject instance-specific prompts to the encoder and task-specific prompts to the decoder. Extensive experiments show that our proposed method is capable of transferring the base codec to various machine tasks and outperforms the competing methods significantly. To our best knowledge, this work is the first attempt to utilize prompting on the low-level image compression task. 

\end{abstract}


\section{Introduction}
End-to-end learned image compression systems~\cite{cheng2020gaussian, guo2021causal, xie2021enhanced} have recently attracted lots of attention due to their competitive compression performance to traditional image coding methods, such as intra coding in VVC~\cite{VVC} and HEVC~\cite{HEVC}. Among them, transformer-based autoencoders~\cite{SwinT-ChARM, TIC, TinyLIC, blockLIC} emerge as attractive alternatives to convolutional neural networks (CNN)-based solutions because of their high content adaptivity. Some even feature lower computational cost than CNN-based autoencoders. In common, most learned image compression systems are designed primarily for \textcolor{black}{human perception}. 


Recently, image coding for machine perception becomes an active research area due to the rising demands for transmitting visual data across devices for high-level recognition tasks. 
Coding techniques in this area mainly include approaches that produce multi-task or single-task bitstreams. The methods with \textcolor{black}{the multi-task bitstream} feature one single compressed bitstream that is able to serve multiple downstream tasks, such as \textcolor{black}{human perception} and machine perception. Most of them~\cite{chamain2021e2emmt, feng2022omnipotent,choi2022sichm, liu2021sssiclrr, yan2021sssic} aim to learn a robust, general image representation via multi-task or contrastive learning. 
However, a general bitstream can hardly be rate-distortion optimal from the perspective of each individual task. 


\begin{figure}[t!]
\centering
\subfigure[Visual Prompt Tuning (VPT)~\cite{VPT}]{
\centering
\includegraphics[scale=0.092,trim=0 990 0 0,clip]{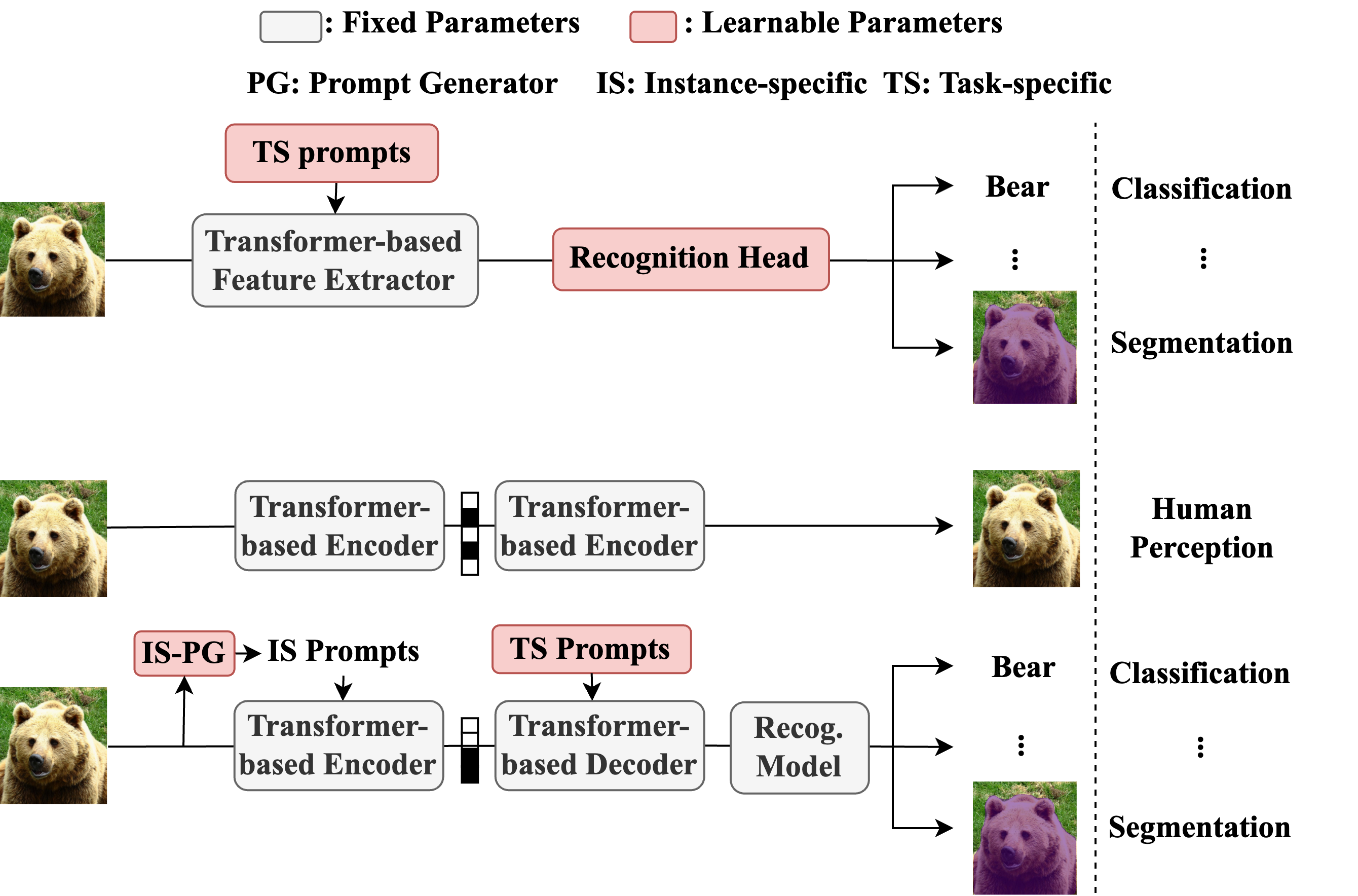}
}
\subfigure[Ours (\textit{TransTIC})]{
\centering
\includegraphics[scale=0.092,trim=0 0 0 860,clip]{Figures/teaser_v6.png}
}
\caption{Comparison of VPT~\cite{VPT} and our proposed TransTIC.}
\vspace{-0.2cm}
\label{fig:teaser}
\end{figure}

The methods with \textcolor{black}{the single-task bitstream} allow the image codec to be tailored for individual downstream tasks, thereby generating multiple task-specific bitstreams. One straightforward approach is to optimize a codec end-to-end for each task~\cite{chamain2021end}. \textcolor{black}{However, given the sheer amount of machine tasks and their recognition networks, along with the ongoing developments of new machine tasks and models, customizing a neural codec, particularly hardware-based, for every one-off machine application would be prohibitively expensive even if not impossible.} Region-of-interest (ROI)-based~\cite{ICCV21} and transferring-based methods~\cite{channelselection} are two preferred solutions. The former performs spatially adaptive coding of images according to an importance map, which can be optimized for different downstream tasks. The latter aims to transfer a pre-trained base codec to a new task without changing the base codec. However, how to transfer efficiently a given codec without re-training is a largely under-explored topic.


In this work, we aim to transfer a well-trained Transformer-based image codec from \textcolor{black}{human perception} to machine perception \textit{without} fine-tuning the codec. Inspired by Visual Prompt Tuning (VPT)~\cite{VPT}, we propose a plug-in mechanism, which injects additional learnable inputs, known as prompts, to the fixed base codec. As shown in Fig.~\ref{fig:teaser}~(a), VPT~\cite{VPT} targets re-using a large-scale, pre-trained Transformer-based feature extractor on different recognition tasks. This is achieved by injecting a small amount of task-specific learnable parameters, prompts, to the Transformer-based feature extractor and learning a task-specific recognition head. 
Different from VPT~\cite{VPT}, which considers only the performance of the downstream recognition task, our task \textcolor{black}{focuses} on image compression, which needs to strike a balance between the downstream task performance and the transmission cost (i.e. the bitrate needed to signal the bistream). Fig.~\ref{fig:teaser}~(b) sketches the high-level design of our proposed method. As shown, the Transformer-based encoder and decoder are initially optimized for \textcolor{black}{human perception} while the recognition model is an off-the-shelf recognition network. To transfer the codec from \textcolor{black}{human perception} to machine perception, we inject prompts to both the encoder and decoder. On the encoder side, we introduce an instance-specific prompt generator to generate instance-specific prompts by observing the input image. On the decoder side, the input image is not accessible. We thus introduce task-specific prompts to the decoder.


Our main contributions are four-fold:
\begin{itemize}[]
\setlength\itemsep{-.3em}
\item Without fine-tuning the codec, we transfer a well-trained Transformer-based image codec from \textcolor{black}{human perception} to machine perception by injecting instance-specific prompts to the encoder and task-specific prompts to the decoder.
\item To the best of our knowledge, this work is the first attempt to utilize prompting techniques on the low-level image compression task. 
\item The plug-in nature of our method makes it easy to integrate any other Transformer-based image codec.
\item \textcolor{black}{Our proposed method is capable of transferring the codec to various machine tasks. Extensive experiments show that our method achieves better rate-accuracy performance than the other transferring-based methods on complex machine tasks.} 
\end{itemize}


\section{Related Work}
\label{sec:related}

\subsection{Learned Image Compression}
\label{Transformer-based codec}

End-to-end learned image compression systems have recently attracted lots of attention due to their competitive compression performance to traditional codecs, such as VVC~\cite{VVC} and HEVC~\cite{HEVC}. Most of them adopt CNN-based autoencoders~\cite{balle2018hyperprior, chen2021context, cheng2020gaussian, hu2020coarsetofine, minnen2018hyperprior} with a hyperprior entropy model. Due to the great success of Transformers in high-level vision tasks, some studies~\cite{SwinT-ChARM, TIC} start to explore their application to low-level vision tasks, such as image compression. In~\cite{SwinT-ChARM}, Zhu~\etal~\cite{SwinT-ChARM} construct an autoencoder using Swin-Transformers~\cite{Swin}, achieving comparable compression performance to the state-of-the-art CNN-based solutions with lower computational complexity. Lu~\etal~\cite{TIC} improves on~\cite{SwinT-ChARM} by replacing patch merging/splitting in Swin-Transformers~\cite{Swin} with the more general convolutional layers. Lu~\etal~\cite{TinyLIC} further adopt Neighborhood Attention Transformers~\cite{hassani2022neighborhood} in place of Swin-Transformers for their more efficient sliding-window attention mechanism. Transformers also find applications in building efficient entropy coding models~\cite{Contextformer, MIMT,VCT}. 

\subsection{Compression for Machine Perception}
\label{Coding for machine}

The success of neural networks in both high-level and low-level vision tasks opens up a new research area known as compression for machine perception. That is, the compressed image features or the decoded image should be suitable for the downstream recognition tasks.
We divide recent works in this emerging research area into three categories according to the characteristics of their coded bitstreams, namely, \textcolor{black}{mutli-task bitstreams}, scalable bitstreams and \textcolor{black}{single-task bitstreams}.

\vspace{-1em}
\paragraph{\textcolor{black}{Mutli-task bitstreams.}} The methods~\cite{chamain2021e2emmt, feng2022omnipotent, le2021icme2e} in this category generate one single compressed bitstream to serve the needs of multiple downstream tasks, such as \textcolor{black}{human perception} and machine perception. 
A straightforward approach is to train an image codec through multi-task learning~\cite{le2021icme2e,chamain2021e2emmt}. More recently, Feng \etal~\cite{feng2022omnipotent} utilize contrastive learning to learn a general image representation for various downstream vision tasks. An obvious disadvantage of these methods is that a single, multi-purpose bitstream can hardly be rate-distortion optimal for individual downstream tasks. Whether such a bitstream can generalize to unseen tasks is an open issue. 



\vspace{-1em}
\paragraph{Scalable bitsreams.} There are also approaches~\cite{choi2022sichm, liu2021sssiclrr, yan2021sssic} that aim to generate a scalable bitstream, which can be partially decoded for simpler machine tasks or fully decoded to reconstruct the input image for \textcolor{black}{human perception}. However, how to arrange potentially multiple image representations for various tasks in a layered manner without introducing redundancy is challenging. 



\vspace{-1em}
\paragraph{\textcolor{black}{Single-task bitstreams.}} Different from the previous two approaches, the \textcolor{black}{methods that produce single-task bitstreams} allow the codec to be tailored for each individual task. A common approach is to fine-tune a pre-trained codec for a specific downstream task~\cite{liu2021cdfmvt, mei2021vaebridge}. \textcolor{black}{However, each task requires a separate model and a machine recognition task normally has a variety of recognition networks to choose from. Customizing a neural codec for each possible choice can be prohibitively expensive, particularly when the neural codec is implemented on specific hardware accelerators.} In comparison, region-of-interest (ROI) coding presents a versatile coding solution. 
For example, Song~\etal~\cite{ICCV21} propose an image codec capable of encoding an input image in a spatially adaptive way according to an importance map. The importance map can be determined at inference time to optimize the decoded image for different uses, e.g. spatial bit allocation or machine perception. 
Another interesting approach is presented in ~\cite{channelselection}, which introduces a learnable task-specific gate module to channel-wisely selects image latents produced by a pre-trained codec for compression. 
Along a similar line of thinking, we propose a novel idea of transferring a pre-trained base codec to different tasks. This is achieved by introducing additional task-specific modules to adapt the base codec without changing its network weights. Specifically, we utilize prompting techniques to transfer a Transformer-based image codec from \textcolor{black}{human perception} to machine perception.  

\subsection{Prompt Tuning}
\label{Prompt finetuning}

The idea of prompting~\cite{lester2021pept, li2021ocpg, liu2021prompt} was first brought up in the field of neural language processing. While transformer-based language models have been a huge success in many language tasks, fine-tuning a pre-trained, large-scale transformer model for a specific downstream task requires huge effort. Prompt tuning offers an attractive alternative, which modifies the input of text encoders while keeping their backbones untouched. Jia \etal~\cite{VPT} are the first to extend this approach to computer vision tasks via inserting learnable task-specific prompts to the input of the vision transformer layers. With only a small number of trainable parameters, it manages to achieve comparable or even superior performance to full fine-tuning in downstream recognition tasks. 
\section{Proposed Method}
\begin{figure*}[t!]
  \centering
  \includegraphics[width=0.99\linewidth]{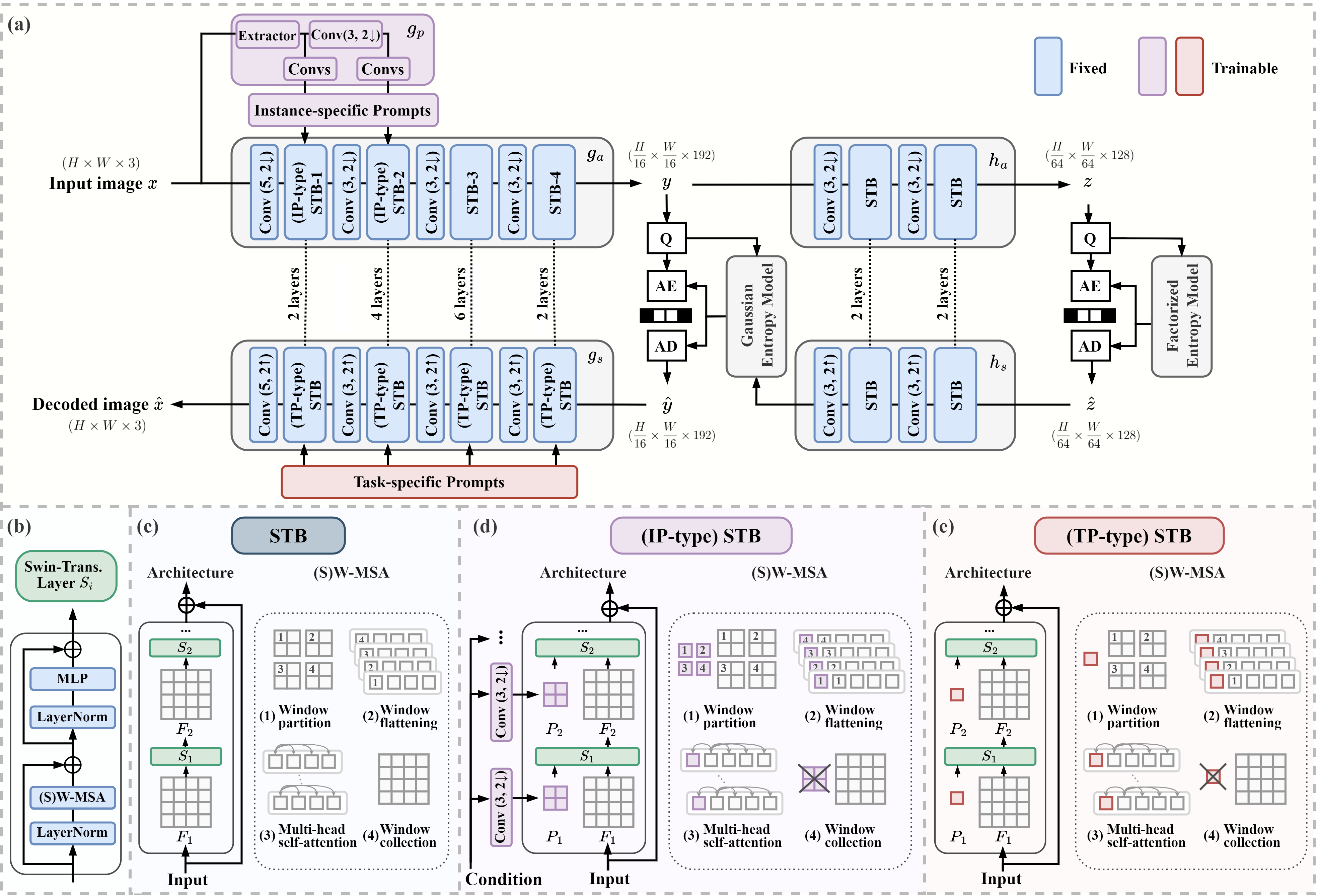}
  \caption{Overall architecture of our proposed method \textit{TransTIC} and the detailed design of STBs.
  }
  \label{fig:framework}
\end{figure*}
\label{sec:method}
Given a Transformer-based image codec well-trained for \textcolor{black}{human perception}, i.e., the image reconstruction task, this work aims for \textcolor{black}{transferring} the codec to achieve image compression for machine perception (e.g., object detection) \textit{without} fine-tuning the codec. We draw inspiration from~\cite{VPT} to propose a plug-in mechanism that utilizes prompting techniques to \textcolor{black}{transfer} the codec. However, unlike~\cite{VPT}, which addresses how to adapt a pre-trained large-scale Transformer-based recognition model to different recognition tasks, our work focuses on \textcolor{black}{transferring} a pre-trained Transformer-based image compression model to tailor image compression for recognition tasks, an application also known as image compression for machines. As such, our performance metric involves not only the recognition accuracy but also the bit rate (in terms of bits-per-pixel) needed to signal the compressed image. In our case, the downstream recognition model can be Transformer-based or convolutional neural network-based. 

In Section~\ref{sec:Preliminary}, we first address a classic paradigm of end-to-end learned image compression. In Section~\ref{sec:Overview}, we outline our proposed framework. This is followed by details of our transferring mechanism in Section~\ref{sec:prompt} and the training objective in Section~\ref{sec:loss}.

\subsection{Preliminaries}
\label{sec:Preliminary}
An end-to-end learned image compression system usually has two major components: a main autoencoder and a hyperprior autoencoder. The main autoencoder consists of an analysis transform ($g_a$ in Fig.~\ref{fig:framework}) and a synthesis transform ($g_s$ in Fig.~\ref{fig:framework}). The analysis transform $g_a$ encodes an RGB image $x \in \mathbb{R}^{H \times W\times 3}$ of height $H$ and width $W$ into its latent representation $y \in \mathbb{R}^{\frac{H}{16} \times \frac{W}{16} \times 192}$ by an encoding distribution $q_{g_a}(y|x)$. The latent $y$ is then uniformly quantized as $\hat{y}$ and entropy encoded into a bitstream by a learned prior distribution $p(\hat{y})$. On the decoder side, $\hat{y}$ is entropy decoded and reconstructed as $\hat{x} \in \mathbb{R}^{H \times W\times 3}$ through a decoding distribution $q_{g_s}(\hat{x}|\hat{y})$ realized by the synthesis transform $g_s$. In the process, the prior distribution $p(\hat{y})$ crucially affects the number of bits needed to be signal the quantized latent $\hat{y}$. It is thus modelled in a content-adaptive manner by a hyperprior autoencoder~\cite{balle2018hyperprior}, which comprises a hyperprior analysis transform ($h_a$ in Fig.~\ref{fig:framework}) and a hyperprior synthesis transform ($h_s$ in Fig.~\ref{fig:framework}). As illustrated, $h_a$ converts the image latent $y$ into the side information $z \in \mathbb{R}^{\frac{H}{64} \times \frac{W}{64} \times 128}$, representing typically a tiny portion of the compressed bitstream. Its quantized version is decoded from the bitstream through $h_s$ to arrive at $p(\hat{y})$. Notably, this work considers a particular implementation of the main and hyperprior autoencoders, the backbones of which are Transformer-based. 


\subsection{System Overview}
\label{sec:Overview}
Fig.~\ref{fig:framework} illustrates our \textcolor{black}{transferable} Transformer-based image compression framework, \textcolor{black}{termed \textit{TransTIC}}. It is built upon~\cite{TIC}, except that the context prior model is replaced with a simpler Gaussian prior for entropy coding. As shown, the main autoencoder $g_a, g_s$ and the hyperprior autoencoder $h_a, h_s$ include Swin-Transformer blocks (STB) as the basic building blocks. These STB are interwoven with convolutional layers to adapt feature resolution in the data pipeline. In this work, the main and hyperprior autoencoders are pre-trained for \textcolor{black}{human perception} (i.e.~the image reconstruction task) and their network weights are fixed during the transferring process. 

To transfer $g_a, g_s$ such that the decoded image $\hat{x}$ is suitable for machine perception, we inject (1) instance-specific prompts produced by $g_p$ into the first two STBs in $g_a$ and (2) task-specific prompts into all the STBs in $g_s$. Section~\ref{sec:prompt} details how these additional prompts are input to these STBs. We note that the prompt generator $g_p$ and the task-specific prompts input to the decoder are learnable and updated according to the machine perception task. That is, the network weights of $g_p$ are task-specific. However, the prompts produced by $g_p$ are instance-specific because they are dependent on the input image. Section~\ref{sec:ablation} presents ablation experiments to justify our design choices. 



\subsection{Prompting Swin-Transformer Blocks}
\label{sec:prompt}
\paragraph{Swin-Transformer blocks (STB).} STB is at the very core of our design. Fig.~\ref{fig:framework}~(c) details its data processing pipeline. It consists of multiple (e.g.~$M$) Swin-Transformer layers (Fig.~\ref{fig:framework}~(b)), with the odd-numbered layers implementing window-based multi-head self-attention (W-MSA) and the even-numbered layers realizing shifted W-MSA (SW-MSA) to facilitate cross-window information exchange. In a Swin-Transformer layer, the input is a set of tokens, each representing a feature vector at a specific spatial location. As shown, the operation of W-MSA (or SW-MSA) has four sequential steps: (1) \textit{window partition} that divides evenly an input feature map \textcolor{black}{$F_i\in\mathbb{R}^{h_i\times w_i\times c_i},i=1,2,\ldots, M$}, in the $i$-th Swin-Transformer layer into non-overlapping windows, (2) \textit{window flattening} that flattens the feature map along the token dimension in each window, (3) \textit{multi-head self-attention} that adaptively updates tokens in each window through self-attending to tokens of the same window, and (4) \textit{window collection} that unflattens the updated tokens in each window and collect tokens from all the windows to form an updated feature map of dimension the same as $F_i$. In symbols, the self-attention in a window for a specific head is given by
\begin{align}
    \text{Attention}(Q,K,V) &= \text{Softmax}(QK^\top/\sqrt{d}+B)V,
    \label{eq:attention}
\end{align}
where $Q = F W_Q$, $K = F W_K$, $V = F W_V$, and $F\in\mathbb{R}^{N \times d}$ is the flattened feature map with $N$ denoting the number of tokens in the window \textcolor{black}{and $d$ the dimension of each token}. 
$W_Q, W_K, W_V\in\mathbb{R}^{d\times d}$ are learnable matrices projecting the input $F$ into query $Q\in\mathbb{R}^{N\times d}$, key $K\in\mathbb{R}^{N\times d}$ and value $V\in\mathbb{R}^{N\times d}$. $B \in \mathbb{R}^{N\times N}$ is a learnable relative position bias matrix. 
\vspace{-1em}
\paragraph{Transferring Encoding STBs via Instance-specific Prompting.}
 To transfer $g_a, g_s$ to machine perception, we propose to inject additional learnable tokens, known as prompts, into the STBs. They interact with the corresponding input tokens in the pre-trained STBs to adapt the encoding process and the decoded image, in order to achieve better rate-accuracy performance for machine perception.  There are two types of prompts: instance-specific prompts and task-specific prompts. Instance-specific prompts are dependent on the input image, while task-specific prompts are task dependent yet invariant to the input image. As shown in Fig.~\ref{fig:framework}~(a), on the encoder side, we introduce \textcolor{black}{an instance}-specific prompt generator $g_p$ that generates instance-specific prompts for the first two STBs (which are referred to as IP-type STBs) based on the input image. $g_p$ itself is task-specific because its network weights are trained for a specific downstream machine task. \textcolor{black}{The network details of $g_p$ are provided in the supplementary document.} Fig.~\ref{fig:framework}~(d) depicts the inner workings of IP-type STBs. They operate similarly to the ordinary STB without prompting, except that additional and separate prompts, denoted collectively as \textcolor{black}{$P_i\in\mathbb{R}^{\frac{h_i}{4}\times \frac{w_i}{4}\times c_i}$}, are introduced in the $i$-th Swin-Transformer layer. In particular, $P_i$ for a specific layer has a spatial resolution that is a quarter of that of $F_i$. This design choice is meant to strike a balance between compression performance and complexity. Morover, $P_i$ is partitioned and flattened in the same way as the image feature $F_i$. In the self-attention step for a specific window and head, the prompts of the same window update the image tokens with the query $Q$, key $K$ and value $V$ matrices in Eq.~\eqref{eq:attention} augmented as follows: 
\begin{equation}
    \label{eq:qkv}
    \begin{aligned}
         Q &= F W_Q, \\
         K &= [F; P]W_K, \\
         V &= [F; P]W_V, \\
    \end{aligned}
\end{equation}
where \textcolor{black}{$P\in\mathbb{R}^{\frac{N}{4} \times d}$ refers collectively to the prompts in the same window as the image tokens $F$} and $[\cdot;\cdot]$ indicates concatenation along the token dimension. With the same $W_Q, W_K, W_V$ as for Eq.~\eqref{eq:attention}, we have
$Q\in\mathbb{R}^{N\times d}$, $K\in\mathbb{R}^{(N+\frac{N}{4})\times d}$ and $V\in\mathbb{R}^{(N+\frac{N}{4})\times d}$. In the window collection step, only image tokens are collected while prompt tokens are discarded. 

\vspace{-1em}
\paragraph{Transferring Decoding STBs via Task-specific Prompting.}
Similarly, we introduce prompts to the STBs in the decoder. Unlike the encoder, the decoder adopts task-specific prompts because the input image is unavailable on the decoder side and communicating instance-specific prompts to the decoder incurs extra overhead. Specifically, these task-specific prompts are input to every STB (referred to as TP-type STB) in the decoder. Fig.~\ref{fig:framework}~(e) illustrates the operation of TP-type STBs. Similar to the IP-type STBs in the encoder, the TP-type STBs are prompted with separate tokens $P_i\in\mathbb{R}^{\frac{N}{4} \times c}$ in different Swin-Transformer layers. However, within a Swin-Transformer layer, the same prompts are shared across fixed-size windows for window-based multi-head self-attention (see window flattening and multi-head self-attention). In other words, in Eq.~\eqref{eq:qkv} when applied to the decoder, $P$ is set to $P_i$ for different windows. This is limited by the fact that the number of fixed-size windows is variable depending on the image size. Training window- and task-specific prompts requires learning a variable number of prompts, which is infeasible.  




\subsection{Training Objective}
\label{sec:loss}
In Fig.~\ref{fig:framework}, the prompt generator network $g_p$ and the task-specific prompts on the decoder side are learnable while the base codec (i.e. $g_a,g_s,h_a,h_s$) is fixed. We construct the training objective in the same way as learning an image compressor. That is, the training involves minimizing a rate-distortion cost
\begin{align}
\label{eq:RDcost}
    \mathcal{L}= \underbrace{-\log p(\hat{z}) -\log p(\hat{y}|\hat{z})}_R 
     + \lambda \underbrace{d(x,\hat{x})}_D,
\end{align}
where $R$ is the estimated rate needed to signal the quantized latent $\hat{y}$ and side information $\hat{z}$, $D$ characterizes the distortion between the input image $x$ and its reconstruction $\hat{x}$, and $\lambda$ is a hyper-parameter. For the sake of machine perception, the distortion measure is perceptual loss, which is evaluated with a recognition network depending on the downstream task. More details can be found in the supplementary document.

\section{Experimental Results}
\label{sec:experiment}
\begin{figure*}[t]
\centering
\subfigure[Classification]{
\centering
\includegraphics[width=0.322\linewidth]{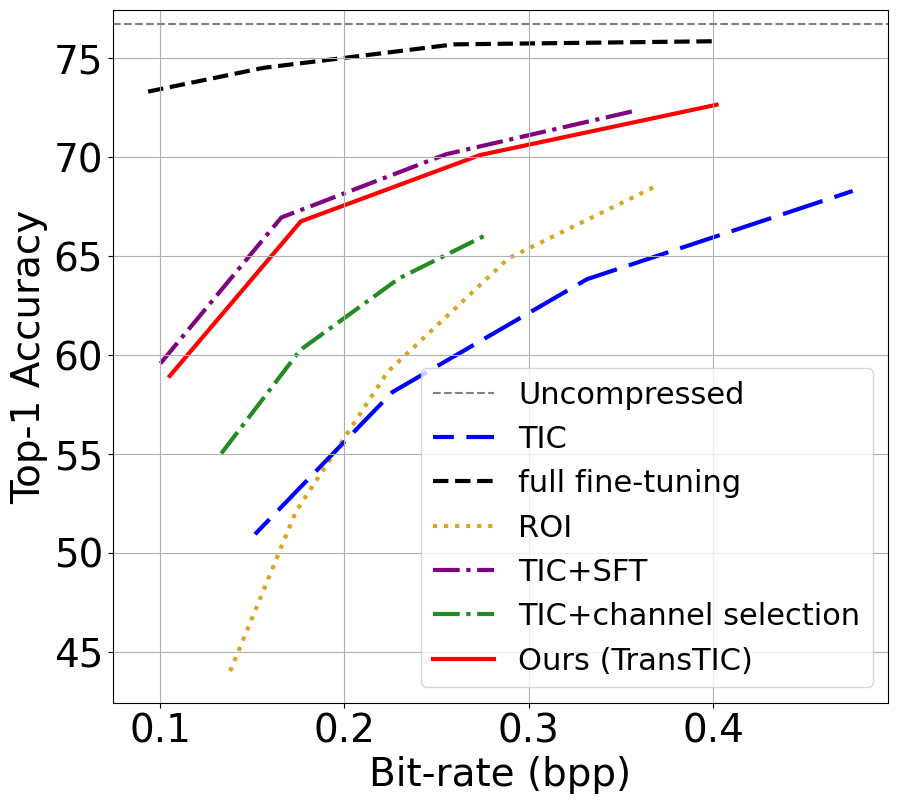}
}
\hspace{-0.35cm}
\subfigure[Object Detection]{
\centering
\includegraphics[width=0.344\linewidth]{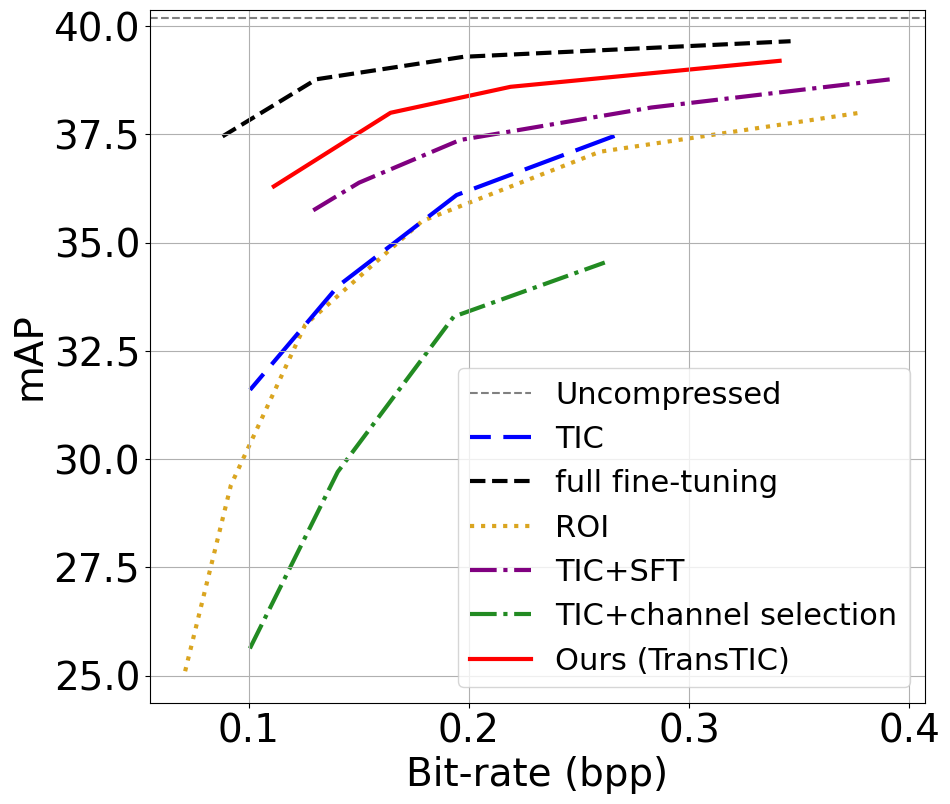}
}
\hspace{-0.35cm}
\subfigure[Instance Segmentation]{
\centering
\includegraphics[width=0.322\linewidth]{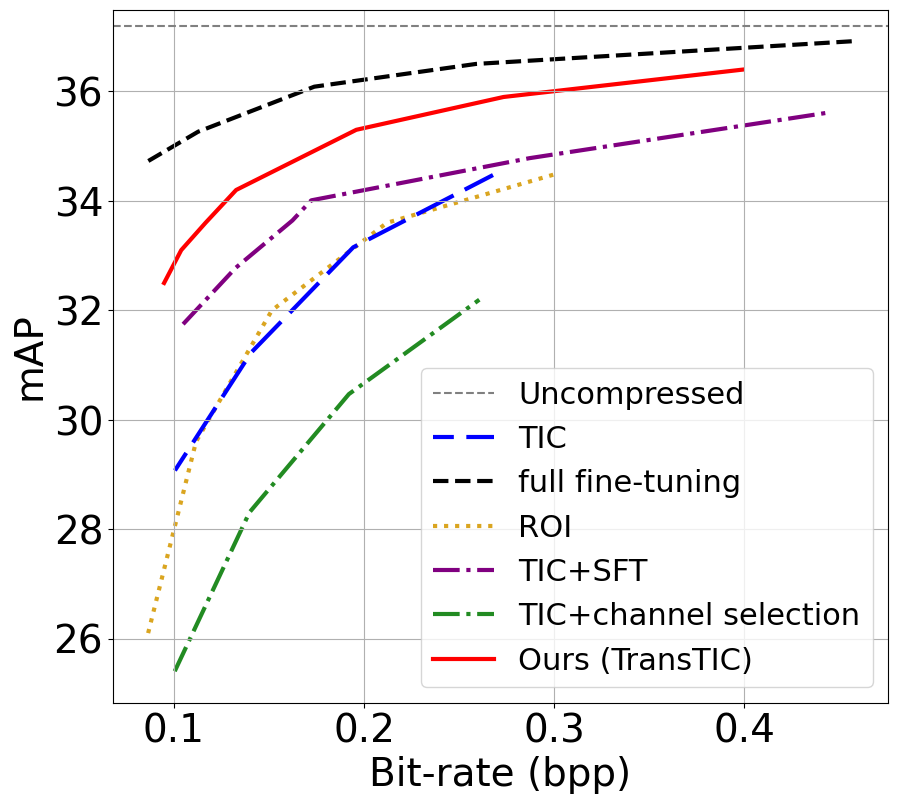}
}
\caption{Rate-accuracy performance comparison under different machine tasks.}
\label{fig:RD}
\end{figure*}
\begin{table*}[]
\caption{BD-Rate, BD-accuracy, and BD-mAP comparison under different machine tasks with \textit{TIC} as anchor.}
\fontsize{8}{8}\selectfont
\centering
\begin{tabular}{c|cc|cc|cc}
\toprule
                        & \multicolumn{2}{c|}{\textbf{Classification}} & \multicolumn{2}{c|}{\textbf{Detection}} & \multicolumn{2}{c}{\textbf{Segmentation}} \\  \cmidrule{2-7}
                                & BD-Rate (\%) $\downarrow$        & BD-accuracy $\uparrow$        & BD-Rate (\%) $\downarrow$        & BD-mAP $\uparrow$        & BD-Rate (\%) $\downarrow$         & BD-mAP $\uparrow$         \\ \midrule
\textit{full fine-tuning}                              &    -     &  16.24  &    -     &  4.11  &    -     &  3.86   \\ \midrule
\textcolor{black}{ROI~\cite{ICCV21}}                                      &  -8.67   &  1.14   &  2.43    &  -0.2  &  -1.45   &  0.04   \\ \midrule
\textit{TIC+SFT}~\cite{wang2018sftgan}                 &  -62.10  &  11.07  &  -26.56  &  1.37  &  -26.47  &  1.4    \\
\textit{TIC+channel selection}~\cite{channelselection} &  -31.13  &  5.88   &  53.66   &  -3.8  &  54.37   &  -2.8   \\
Ours (\textit{TransTIC})                               &  -58.77  &  10.02  &  -46.07  &  2.72  &  -45.83  &  2.66   \\
\bottomrule
\end{tabular}
\label{table:main_RD}
\end{table*}

\paragraph{Training Details and Datasets.}
We evaluate our method on three machine tasks: classification, object detection, and instance segmentation. We first follow~\cite{TIC} to train the base codec composed of $g_a, g_s, h_a, h_s$ on Flicker2W~\cite{liu2020unified} for \textcolor{black}{human perception}. In this case, the distortion measure $d(\cdot,\cdot)$ in Eq.~\eqref{eq:RDcost} is mean-squared error, and $\lambda$ is chosen to be 0.0018, 0.0035, 0.0067, and 0.013 to arrive at four separate codecs for variable-rate compression. By freezing these pre-trained base codecs, we then train the prompt generator $g_p$ together with the task-specific decoder-side prompts on ImageNet-\textit{train}~\cite{deng2009imagenet} for classification and on COCO2017-\textit{train}~\cite{cocodataset} for object detection and instance segmentation. In the process, $d(\cdot,\cdot)$ in Eq.~\eqref{eq:RDcost} is evaluated based on the perceptual loss using a pre-trained ResNet50~\cite{he2016deep}, Faster R-CNN~\cite{shaoqing2015faster} and Mask R-CNN~\cite{kaiming2017Mask} for classification, object detection and instance segmentation, respectively.

\vspace{-1em}
\paragraph{Evaluation.}
For classification, we use ImageNet-\textit{val}~\cite{deng2009imagenet} as the test set and a pre-trained ResNet50~\cite{he2016deep} as the downstream recognition network. For object detection and instance segmentation, we test the competing methods on COCO2017-\textit{val}~\cite{cocodataset} using a pre-trained Faster R-CNN~\cite{shaoqing2015faster} and Mask R-CNN~\cite{kaiming2017Mask} as the downstream recognition networks, respectively. Note that these recognition networks are the same as those utilized for learning $g_p$ and the decoder-side prompts. We adopt top-1 accuracy as the quality metric for classification and mean average precision (mAP) for both detection and instance segmentation.

\vspace{-1em}
\paragraph{Baseline Methods.}
The baseline methods include: (1) using the base codec ($g_a, g_s, h_a, h_s$) trained for \textcolor{black}{human perception} without prompting, known as~\textit{TIC}, (2) fine-tuning \textit{TIC} end-to-end for the downstream machine tasks, i.e.,~\textit{full fine-tuning}, (3) transferring~\textit{TIC} by introducing spatial feature transform (SFT) layers, termed~\textit{TIC+SFT}, (4) selecting partial channels of the image latent $y$ produced by~\textit{TIC} to perform coding for machine perception, termed~\textit{TIC+channel selection}, and (5) adopting \textcolor{black}{region-of-interest (ROI) coding} proposed in~\cite{ICCV21}, which modulates a CNN-based compression backbone with SFT layers~\cite{wang2018sftgan} according to a ROI map. 

For~\textit{TIC+SFT}, we introduce a SFT layer after every Swin-Transformer block in the encoder and decoder. In particular, the affine parameters for SFT are produced by a task-specific network that takes the coding image as input. In a sense,~\textit{TIC+SFT} presents an alternative to our prompting technique. For~\textit{TIC+channel selection}, we follow~\cite{channelselection} in implementing a task-specific channel selection module, which observes the image latent for channel selection, and a task-specific transform module, which converts the selected latent channels into feature maps suitable for the downstream recognition network. Both~\textit{TIC+SFT} and~\textit{TIC+channel selection} use the same base codec (i.e.,~\textit{TIC}) and training protocol as our~\textit{TransTIC} to train additional task-specific networks for spatial feature transform or channel selection. More details about their implementation are provided in the supplementary document. 

For \textcolor{black}{ROI}, we follow~\cite{ICCV21} to extract the Grad-CAM~\cite{selvaraju2017grad} output based on the pre-trained ResNet50~\cite{he2016deep} as the ROI map for classification. For object detection and instance segmentation, we generate binary ROI maps according to the foreground predictions of Faster R-CNN~\cite{shaoqing2015faster} and Mask R-CNN~\cite{kaiming2017Mask}, respectively.

\subsection{Rate-Accuracy Comparison}
\begin{figure}[t]
\centering
\includegraphics[width=0.78\linewidth]{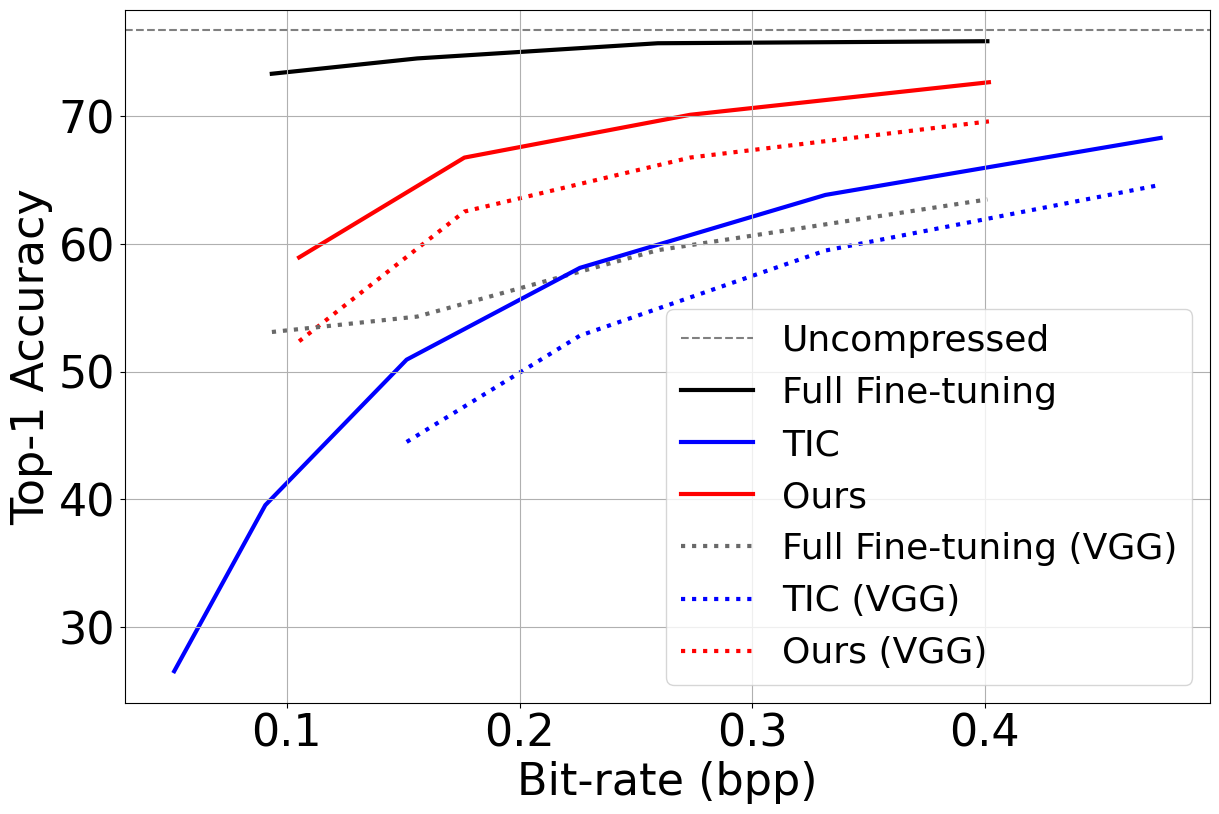}
\caption{Rate-accuracy comparison on classification with two different recognition networks (ResNet50 and VGG19).}
\label{fig:RD_generalization}
\vspace{-0.15cm}
\end{figure}

Fig.~\ref{fig:RD} visualizes the rate-accuracy plots for the competing methods. Table~\ref{table:main_RD} summarizes the average bit-rate savings and accuracy improvements using the Bjontegaard metrics~\cite{Bjntegaard2001CalculationOA}. \textcolor{black}{BD-accuracy/mAP are computed similarly to BD-PSNR with PSNR replaced with the top-1 accuracy or mAP. Negative BD-rate numbers suggest rate saving at the same quality/accuracy level, while positive BD-PSNR/mAP/accuracy numbers suggest quality/accuracy improvements at the same bit rate.}
From Fig.~\ref{fig:RD} and Table~\ref{table:main_RD}, we make the following observations. First, (1) our~\textit{TransTIC} and~\textit{TIC+SFT} outperform~\textit{TIC+channel selection} across all the recognition tasks. This is attributed to the fact that both~\textit{TransTIC} and~\textit{TIC+SFT} are able to achieve spatially adaptive coding (see Section~\ref{sec:qualitative} for their decoded images). Second, (2) our~\textit{TransTIC} performs comparably to~\textit{TIC+SFT} on the classification task and outperforms \textit{TIC+SFT} on more complicated tasks, such as object detection and instance segmentation. This suggests that our prompting technique works more effectively than spatial feature transform~\cite{wang2018sftgan} in terms of transferring our transform-based codec. Third, (3) \textcolor{black}{ROI}~\cite{ICCV21} performs the worst \textcolor{black}{in between methods with spatially adaptive ability (i.e.,~\textit{TransTIC}, \textit{TIC+SFT} and ROI). It relies heavily on the quality of the ROI mask, which can later be seen in Fig.~\ref{fig:visualization}}. Lastly, (4) \textit{full fine-tuning} achieves the best performance as expected. However, this comes at the expense of having to fully fine-tune the codec for the downstream recognition model. \textcolor{black}{With a wide variety of machine tasks and their recognition networks, customizing neural codecs, particularly hardware-based, for each task is impractical. One feasible approach would be to re-purpose existing neural codecs already in mass production for new machine tasks.} Fig.~\ref{fig:RD_generalization} shows that the image codec trained with \textit{full fine-tuning} generalizes poorly to other unseen recognition models. Taking the classification task as an example, we replace the downstream classification model ResNet50, which is used for training under the settings of \textit{full fine-tuning} and our~\textit{TransTIC}, with VGG19 at test time. As shown, the accuracy of~\textit{full fine-tuning} drops severely by 15\%-20\%. In contrast, our \textit{TransTIC} has a less sharp decline of 3\%-7\%. Part of their accuracy loss comes from the smaller model capacity of VGG19 than that of ResNet50. This is evidenced by the \textcolor{black}{4\%-8\%} accuracy loss on~\textit{TIC} when VGG19 is used in place of ResNet50.

\begin{figure}[t]
\centering
\subfigure[Detection]{
    \centering
    \includegraphics[width=.44\textwidth]{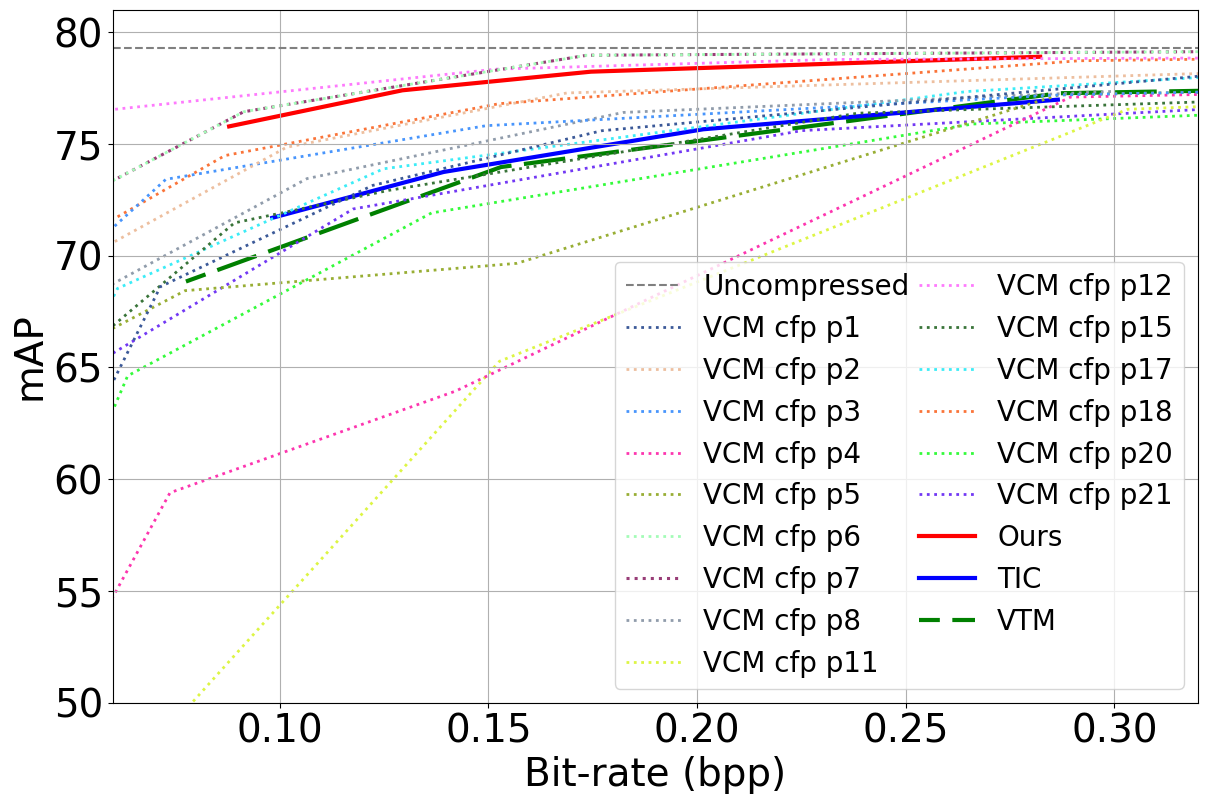}
    \label{fig:rd-a}
    }
\subfigure[Segmentation]{
    \centering
    \includegraphics[width=.44\textwidth]{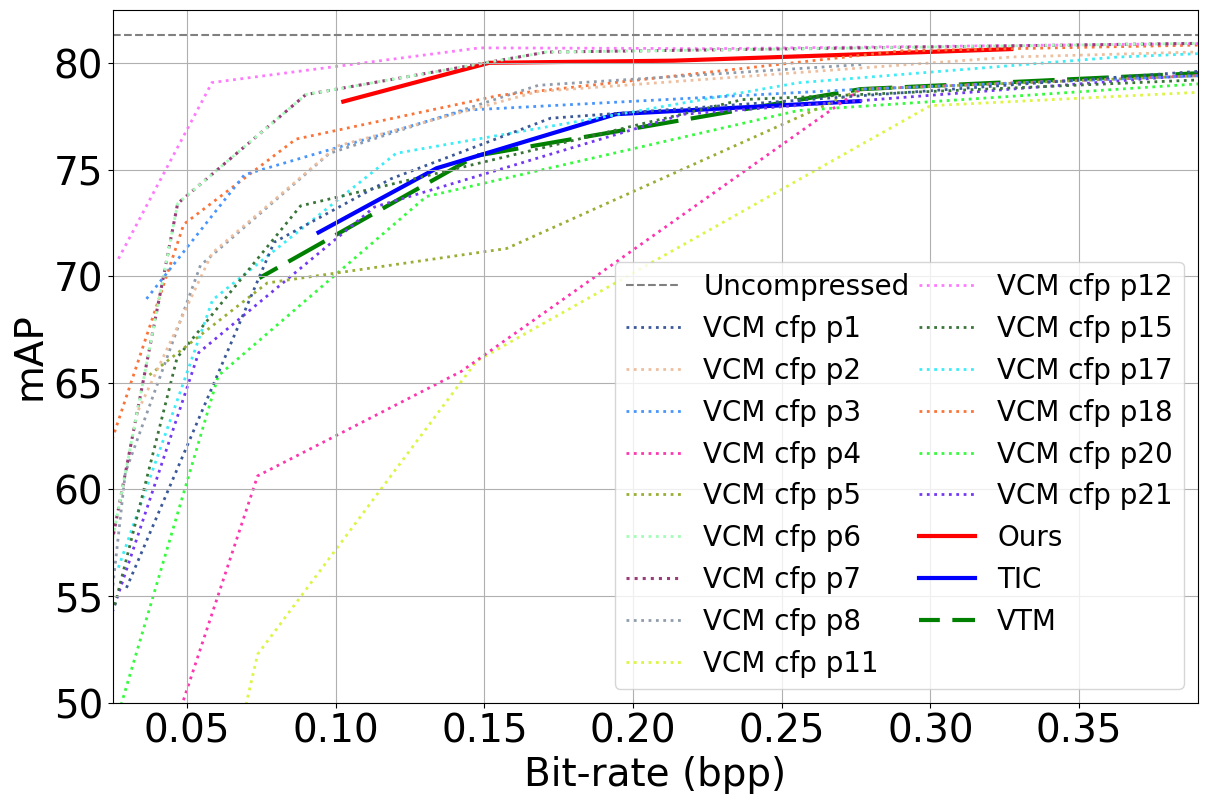}
    \label{fig:rd-b}
}
\caption{Rate-mAP comparison against MPEG VCM proposals on OpenImages v6 dataset~\cite{OpenImages}.}
\label{fig:MPEG}
\vspace{-0.2cm}
\end{figure}

\textcolor{black}{To further validate the performance of our proposed method, we also compare our TransTIC with the methods recently submitted to the call-for-proposals (CFP) competition of the MPEG VCM standard based on their test protocol\cite{MPEGVCMctc}. The results of these competing methods are from the CFP test report~\cite{MPEGm61010}. As shown in Fig.~\ref{fig:MPEG}, our TransTIC performs comparably to the top performers in terms of rate-accuracy performance. However, our base codec has the additional constraint that it is optimized for human perception, while the top performers (e.g. p12, p6, p7) optimize the entire codec end-to-end for machine tasks. This shows the potential of our TransTIC.}

\subsection{Qualitative Results}
\label{sec:qualitative}
\begin{figure*}[t]
\centering
\subfigure[Classification]{
\centering\
\includegraphics[width=0.98\linewidth]{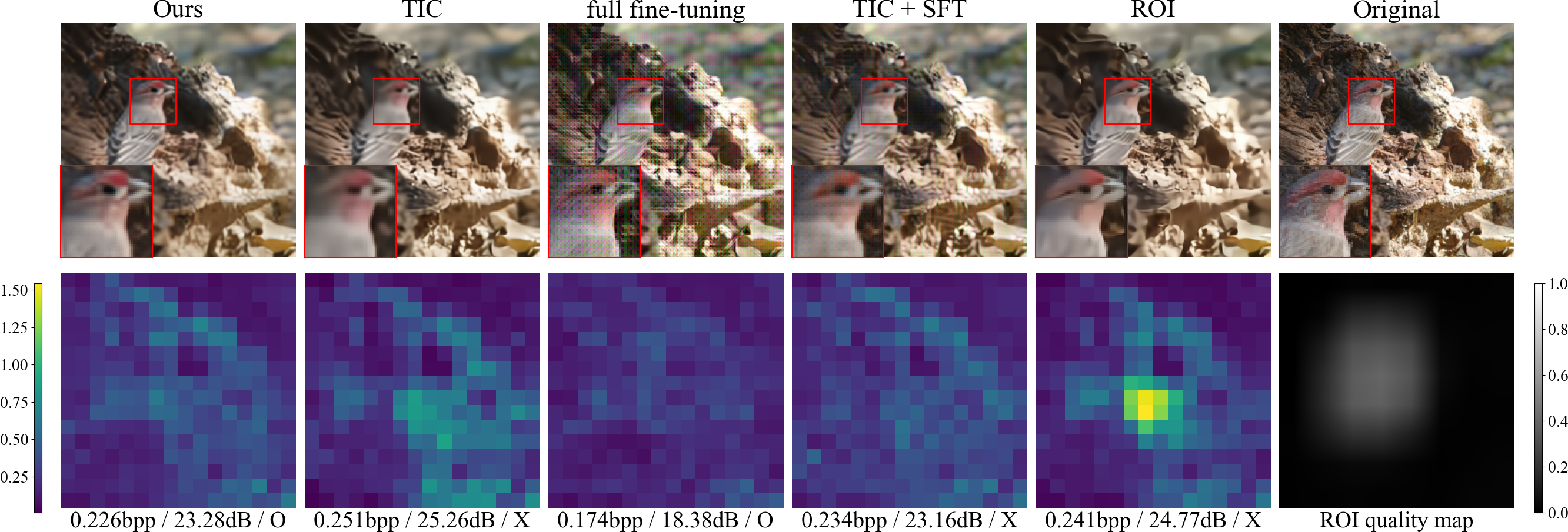}
}
\subfigure[Object Detection]{
\centering
\includegraphics[width=0.98\linewidth]{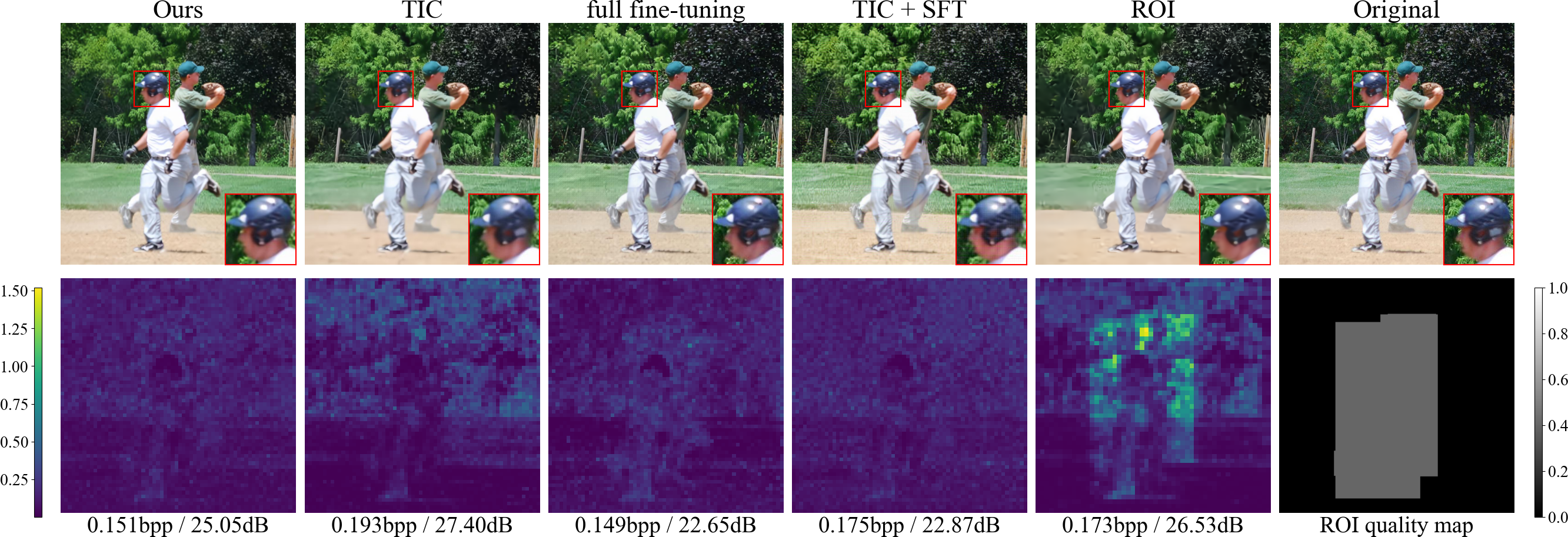}
}
\caption{Visualization of decoded images (first row) and bit allocation map of latent $\hat{y}$ (second row). The rightmost image of the second row shows the quality map used for ROI method. The text below each map denotes the corresponding rate / PSNR / prediction result (classification only). The values of bit allocation maps denote the average negative log likelihood of each element in $\hat{y}$ across all channels.}
\label{fig:visualization}
\end{figure*}

Fig.~\ref{fig:visualization} presents the decoded images and the corresponding bit allocation \textcolor{black}{maps produced by the competing methods. As shown, the base codec \textit{TIC}, which is optimized for \textcolor{black}{human perception}, tends to spend more bits on coding complex regions (e.g.~the rocky surface in Fig.~\ref{fig:visualization}~(a) and the background forest in Fig.~\ref{fig:visualization}~(b)), which may be less relevant to the downstream recognition tasks. In contrast, our TransTIC and the other methods optimized for machine tasks shift more bits from the background to the foreground, resulting in generally more sharper foreground objects.}

\subsection{Complexity Comparison}
\label{sec:complexity}
\textcolor{black}{Table~\ref{tab:complexity} compares the complexity of the competing methods in terms of the kilo-multiply-accumulate-operations per pixel (kMACs/pixel) and model size. Through transferring the pre-trained \textit{TIC}, our \textit{TransTIC} only needs to learn for each task a prompt generation network on the encoder side and several task-specific prompts on the decoder side. The number of these additional parameters is about one fifth of that of a complete \textit{TIC} (7M). 
As compared with \textit{TIC}, the increase in kMACs/pixel (i.e. from 188 to 202) on the decoder side is modest. In contrast, our~\textit{TransTIC} has a much more complex encoder than~\textit{TIC} because of incorporating a prompt generation network, which must adapt the encoder-side prompts to the downstream task and the current input image. Nevertheless, \textit{TransTIC} offers significantly lower kMACs/pixel and parameters than ROI~\cite{ICCV21} and \textit{TIC+SFT}~\cite{wang2018sftgan}, both of which utilize more computationally heavy networks for SFT layers. Furthermore, ROI~\cite{ICCV21} currently uses a pre-trained recognition network to generate the ROI mask. This results in much higher kMACs/pixel on the encoder side. Table~\ref{tab:complexity} provides the kMACs/pixel needed to generate the ROI masks for different tasks. Last but not least, \textit{TIC+channel selection}~\cite{channelselection} has a low-complexity decoder (which is composed of only the hyperprior decoder and the transform module) because it need not reconstruct the input image. However, it performs much worse than our~\textit{TransTIC} (see Fig.~3).}

\begin{table}[]
\centering
\caption{Comparison of the kMACs/pixel and model size.}
\setlength{\tabcolsep}{3.0pt}
\fontsize{8}{8}\selectfont
\begin{tabular}{cc|cc|cc}
\toprule
\multicolumn{2}{l|}{}                                                                             & \multicolumn{2}{c|}{\textbf{kMACs/pixel}} & \multicolumn{2}{c}{\textbf{Params (M)}} \\
\multicolumn{2}{l|}{}                                                                             & Encoder         & Decoder        & Encoder            & Decoder            \\ \midrule 
\multicolumn{2}{c|}{\textit{TIC}}                                                                 & 142.54          & 188.52         & 3.65               & 3.86               \\ \midrule 
\multicolumn{1}{c}{\multirow{4}{*}{ROI~\cite{ICCV21}}}  & Autoencoder only  & 800.36          & 679.80         & 21.91              & 5.65               \\
\multicolumn{1}{c}{}                                                          & Classification    & 882.46          & 679.80         & 47.47              & 5.65               \\
\multicolumn{1}{c}{}                                                          & Detection         & 991.13          & 679.80         & 63.39              & 5.65               \\
\multicolumn{1}{c}{}                                                          & Segmentation      & $\approx$997.85          & 679.80         & 66.03              & 5.65               \\ \midrule 
\multicolumn{2}{c|}{\textit{TIC+SFT}~\cite{wang2018sftgan}}                & 686.39          & 462.23         & 12.08              & 9.62               \\
\multicolumn{2}{c|}{\textit{TIC+channel selection}~\cite{channelselection}} & 142.54          & 25.13          & 3.76               & 1.96               \\
\multicolumn{2}{c|}{Ours (\textit{TransTIC})}                                                  & 332.03          & 202.60         & 5.24               & 3.89               \\  
\bottomrule
\end{tabular}
\label{tab:complexity}
\end{table}

\subsection{Ablation Experiments}
\label{sec:ablation}
\paragraph{IP-type~vs.~TP-type STBs.} This ablation experiment investigates how the prompting type, instance-specific (IP-type) or task-specific (TP-type), in the encoder and decoder STBs may impact the rate-accuracy performance. We explore four variants of the proposed method, as summarized in Table~\ref{tab:ablation_settings}, where the IP-shared-type refers to instance-specific prompting with the same instance-specific prompts shared across windows in a Swin-Transformer layer. When included in the encoder STBs, these shared prompts are learned by introducing a spatially adaptive pooling layer after the prompt generation network. This ensures that the number of prompts is invariant to the image size. 

\begin{table}[]

\caption{Ablation variants of IP-type~vs.~TP-type STBs.}
\fontsize{8}{8}\selectfont
\centering

\begin{tabular}{c|c|c}
\toprule
Variants & Encoder Prompting    & Decoder Prompting\\
\midrule
A      & IP-type         & TP-type      \\
B      & TP-type         & TP-type      \\
C      & IP-type         & Disabled    \\
D      & IP-shared-type       & TP-type      \\
\bottomrule
\end{tabular}
\label{tab:ablation_settings}
\end{table}
\begin{figure}[t]
\centering
\subfigure[Classification]{
\centering
\includegraphics[width=0.49\linewidth]{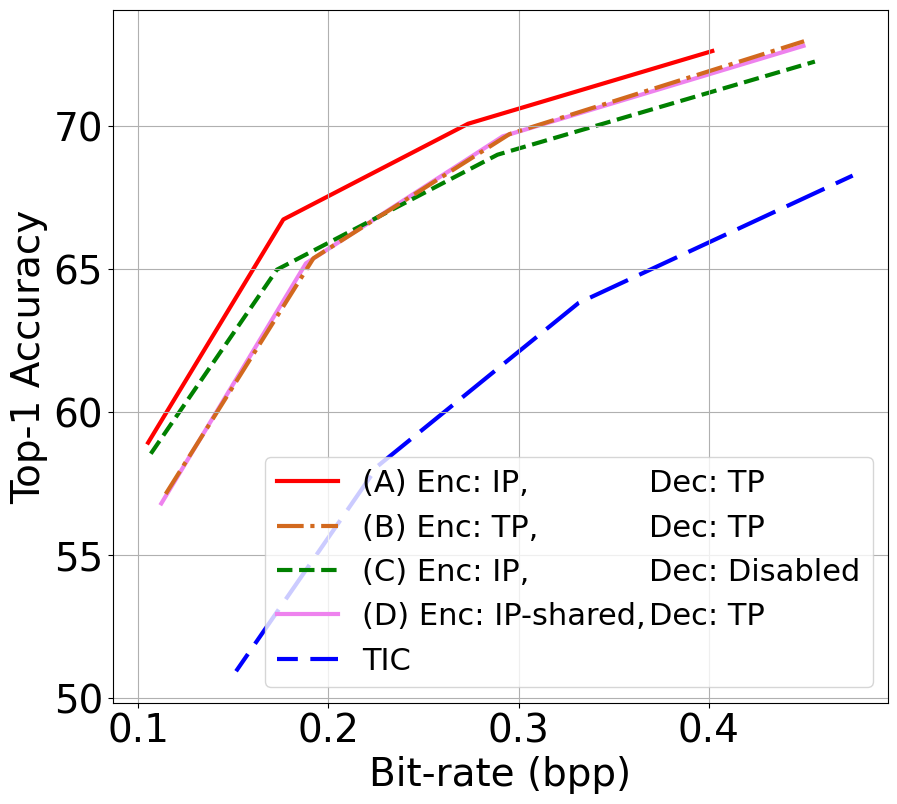}
}
\hspace{-0.35cm}
\subfigure[Object Detection]{
\centering
\includegraphics[width=0.49\linewidth]{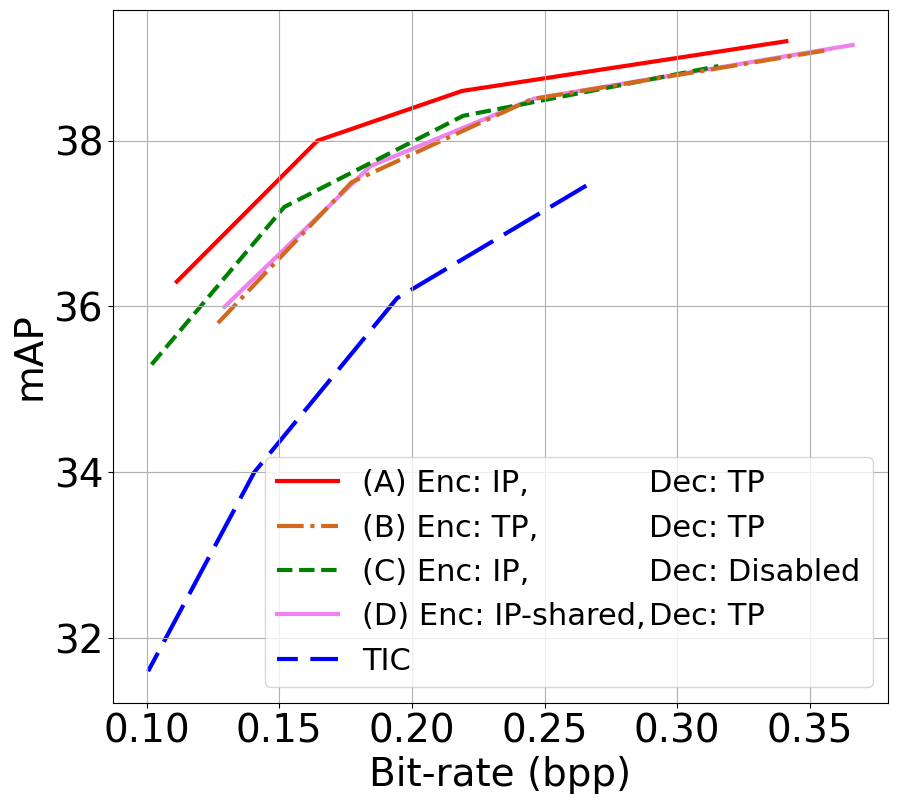}
}
\caption{Ablation on IP-type and TP-type STBs.}
\label{fig:STB_ablation}
\end{figure}

In Fig.~\ref{fig:STB_ablation}, \textcolor{black}{we make the following observations. First,} a comparison of our full model (variant A) and variant B shows that IP-type prompting works more effectively than TP-type prompting on the encoder side. Intuitively, adapting the encoding process according to the input image helps generate a bitstream better suited for the downstream recognition task in the rate-accuracy sense. 
Second, compared to the full model, disabling the decoder-side prompting (variants A vs. C) results in only a moderate accuracy loss, suggesting that the encoder-side prompting is more critical. Third, on the encoder side, IP-shared prompting (variant D) performs worse than IP-type prompting (variant A). That is, spatially adaptive prompting is indispensable. Last but not least, all the variants outshine~\textit{TIC} significantly, implying that prompting is effective in transferring~\textit{TIC} from \textcolor{black}{human perception} to machine perception.   

\begin{figure}[t]
\centering
\subfigure[Classification]{
\centering
\includegraphics[width=0.49\linewidth]{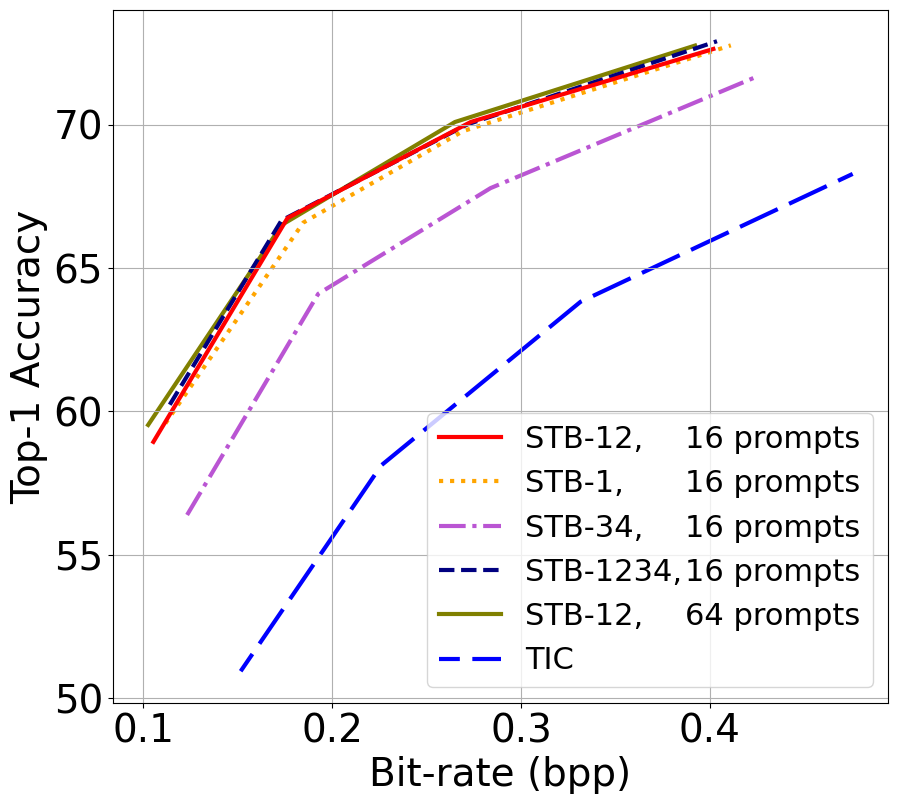}
}
\hspace{-0.35cm}
\subfigure[Object Detection]{
\centering
\includegraphics[width=0.49\linewidth]{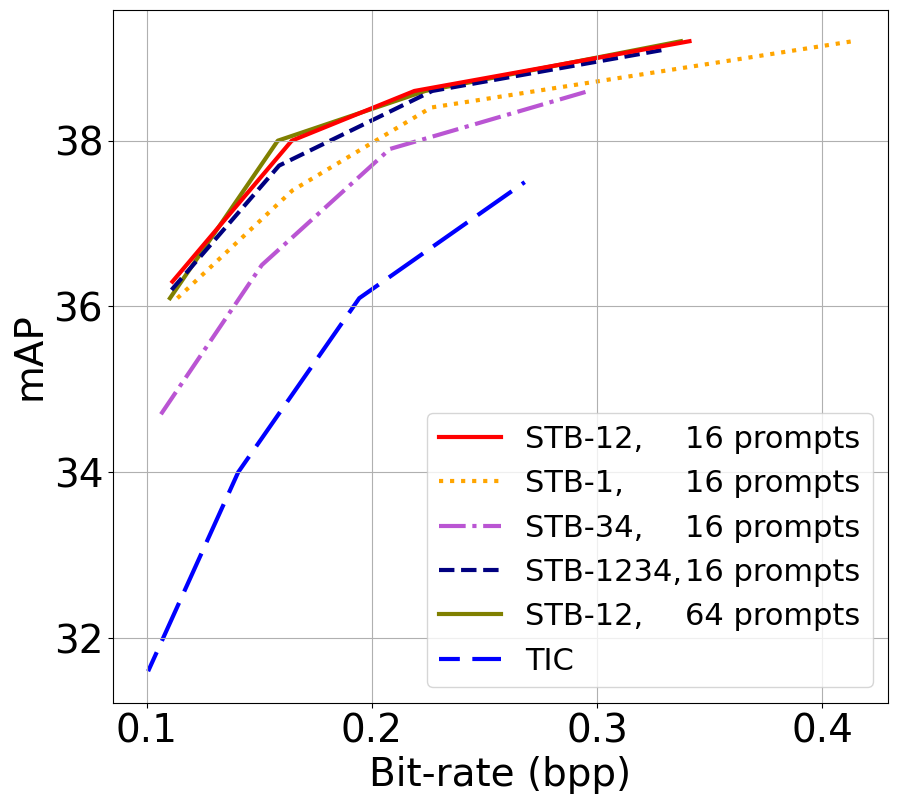}
}
\caption{Ablation on the depth and number of prompts.}
\label{fig:prompts_ablation}
\vspace{-0.2cm}
\end{figure}

\vspace{-1em}
\paragraph{Prompt Depth.} 
Fig.~\ref{fig:prompts_ablation} analyzes which and how many STBs to inject prompts on the encoder side. As shown, injecting prompts to early STBs closer to the input image (e.g.~STB-1 with 16 prompts, and STB-1,2 with 16 prompts) is more effective than injecting them to later STBs (e.g.~STB-3,4 with 16 prompts). Comparing the two early prompting variants (STB-1 with 16 prompts~vs.~STB-1,2 with 16 prompts), it is clear that the performance gap between them is rather limited on the easier classification task, but becomes more significant on object detection. Also, the full injection variant (STB-1,2,3,4 with 16 prompts) performs comparably to the early prompting variant (STB-1,2 with 16 prompts). We thus choose to inject prompts to STB-1,2 only.

\vspace{-1em}
\paragraph{Prompt Numbers.}
Fig.~\ref{fig:prompts_ablation} also ablates different design choices on prompt numbers. 
In our design, the number of prompts in a window is chosen empirically to be 16, which is one quarter of that (i.e.~64) of the image tokens. As shown in Fig.~\ref{fig:prompts_ablation}, increasing the number of prompts from 16 to 64 (cp.~STB-1,2 with 64 prompts vs. STB-1,2 with 16 prompts) brings little benefit on recognition accuracy. We thus stick to 16 prompts to save storage and computational cost.




\section{Conclusion}
This paper utilizes prompting techniques to transfer a well-trained Transformer-based image codec from \textcolor{black}{human perception} to machine perception. Instead of retraining the codec, we introduce additional instance-specific prompts to the Swin-Transformer layers in the encoder and task-specific prompts to the decoder. Experimental results show that our TransTIC achieves comparable or better rate-accuracy performance than the other transferring methods on various machine tasks. 

\section*{Acknowledgement}
{\textcolor{black}{This work is supported by National Science and Technology Council, Taiwan under Grants NSTC 111-2634-F-A49-010- and MOST 110-2221-E-A49-
065-MY3, MediaTek, and National Center for High-performance Computing.}}

{\small
\bibliographystyle{ieee_fullname}
\bibliography{egbib}
}


\beginsupplement

\predate{}
\postdate{}

\title{\textbf{TransTIC: Transferring Transformer-based Image Compression \\ from Human Perception to Machine Perception \\ \textit{Supplementary Materials}}}

\author{
Yi-Hsin Chen \quad Ying-Chieh Weng \quad Chia-Hao Kao \quad Cheng Chien \\ Wei-Chen Chiu \quad Wen-Hsiao Peng \\
National Yang Ming Chiao Tung University, Taiwan\\
\tt\small \{yhchen12101.cs09@, wengyc.cs09@, chiahaok.cs10@, cchien1999@cs.\}nycu.edu.tw \\ \tt\small \{walon, wpeng\}@cs.nctu.edu.tw
}

\date{}

\maketitle
\ificcvfinal\thispagestyle{empty}\fi

This supplementary document provides the following additional materials and results to assist with the understanding of our \textit{TransTIC}:
\begin{itemize}
    \setlength{\itemsep}{0pt}
    \item Implementation details in Section~\ref{sec:implementation};
    \item Rate-accuracy comparison with the state-of-the-art traditional codec VVC in Section~\ref{sec:VVC};
    \item More ablation experiments in Section~\ref{sec:ablation_supp};
    \item More qualitative results in Section~\ref{sec:qualitative_supp}.
\end{itemize}






\section{Implementation Details}
\label{sec:implementation}

\subsection{Perceptual Loss}
\begin{figure}[t]
  \centering
  \includegraphics[width=0.75\linewidth]{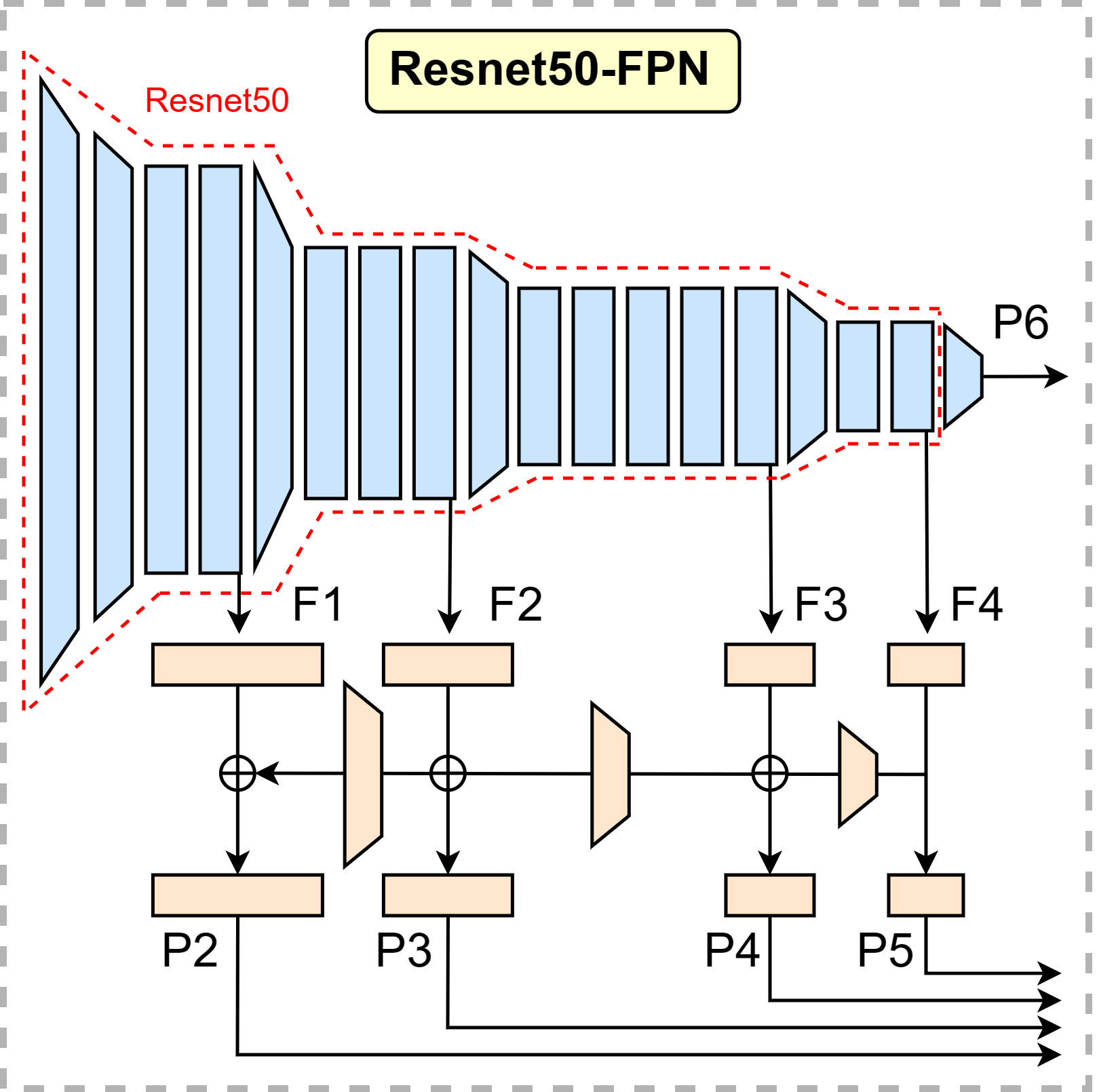}
  \caption{Architecture of Resnet50-based FPN, which shows the features selected for evaluating the perceptual loss.}
  \label{fig:perceptual}
\end{figure}

To train the prompt generator network $g_p$ and the decoder-side prompts for downstream recognition tasks, the distortion measure $d(\cdot,\cdot)$ in Eq.~(3) of the main paper is chosen to be the perceptual loss.
Specifically, the perceptual loss is evaluated based on a pre-trained ResNet50~\cite{he2016deep}, Faster R-CNN~\cite{shaoqing2015faster} and Mask R-CNN~\cite{kaiming2017Mask} for classification, object detection and instance segmentation, respectively. Fig.~\ref{fig:perceptual} illustrates a ResNet50-based Feature Pyramid Network (FPN), which serves as the feature extractor in Faster R-CNN and Mask R-CNN. For the classification task, the perceptual loss is evaluated in the feature space of F1, F2, F3, and F4: 
\begin{align}
\label{eq:cls_perceputual}
    d(x,\hat{x}) = \frac{1}{4}\cdot\sum_{l=1}^4 MSE(F_l(x), F_l(\hat{x})),
\end{align} where $x$ and $\hat{x}$ are the input and decoded images, respectively.
For the tasks of object detection and instance segmentation, the perceptual loss is evaluated in the feature space of P2, P3, P4, P5, and P6:
\begin{align}
\label{eq:det_perceputual}
    d(x,\hat{x}) = \frac{1}{5}\cdot\sum_{l=2}^6 MSE(P_l(x), P_l(\hat{x})).
\end{align}
In Fig.~\ref{fig:perceptual}, the network weights are initialized using a separate pre-trained model, depending on the downstream task.

\begin{figure}[t]
  \centering
  \includegraphics[width=0.99\linewidth]{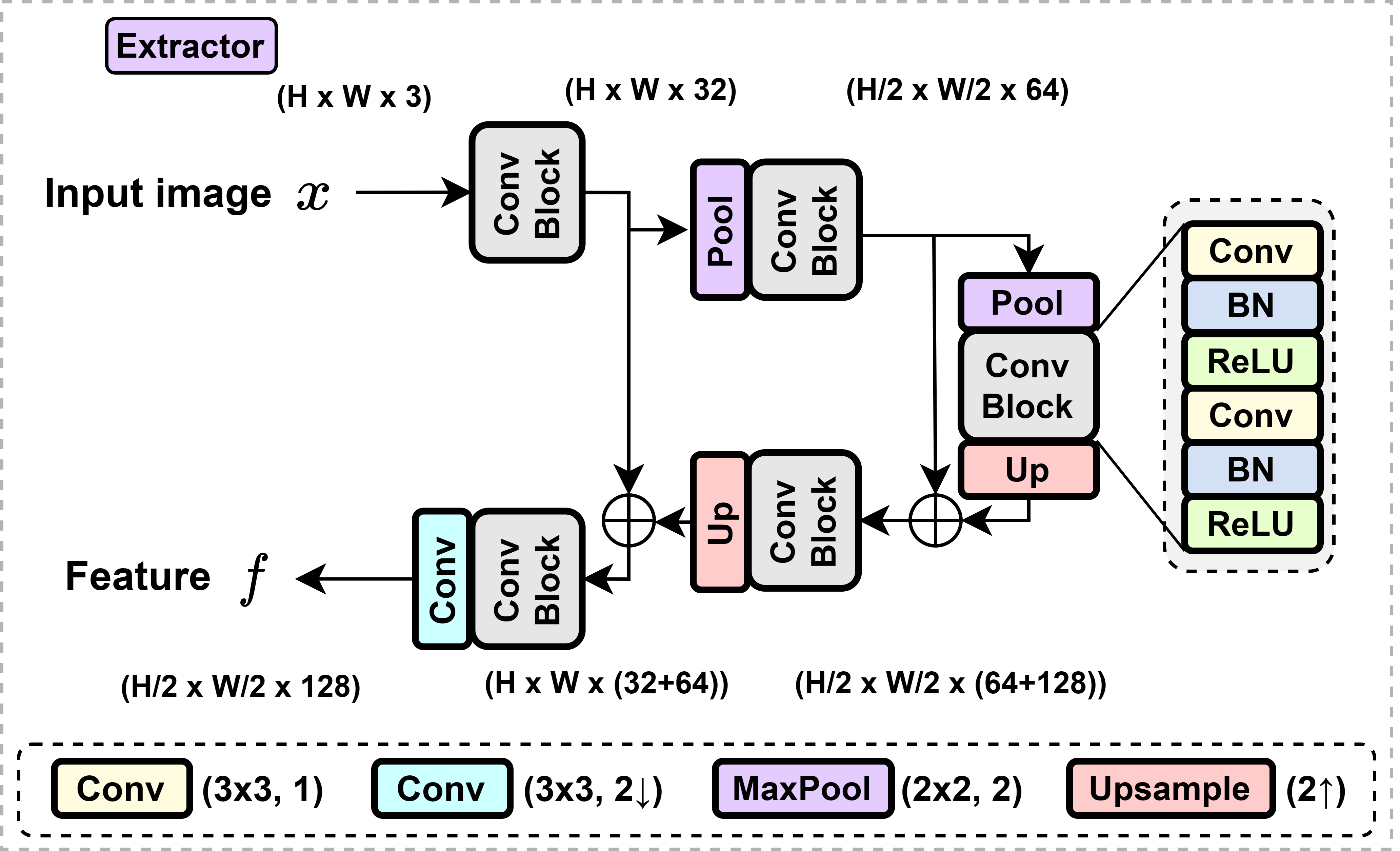}
  \caption{Architecture of the extractor in our prompt generator $g_p$.}
  \label{fig:unet}
\end{figure}

\begin{figure*}[t!]
  \centering
  \includegraphics[width=0.99\linewidth]{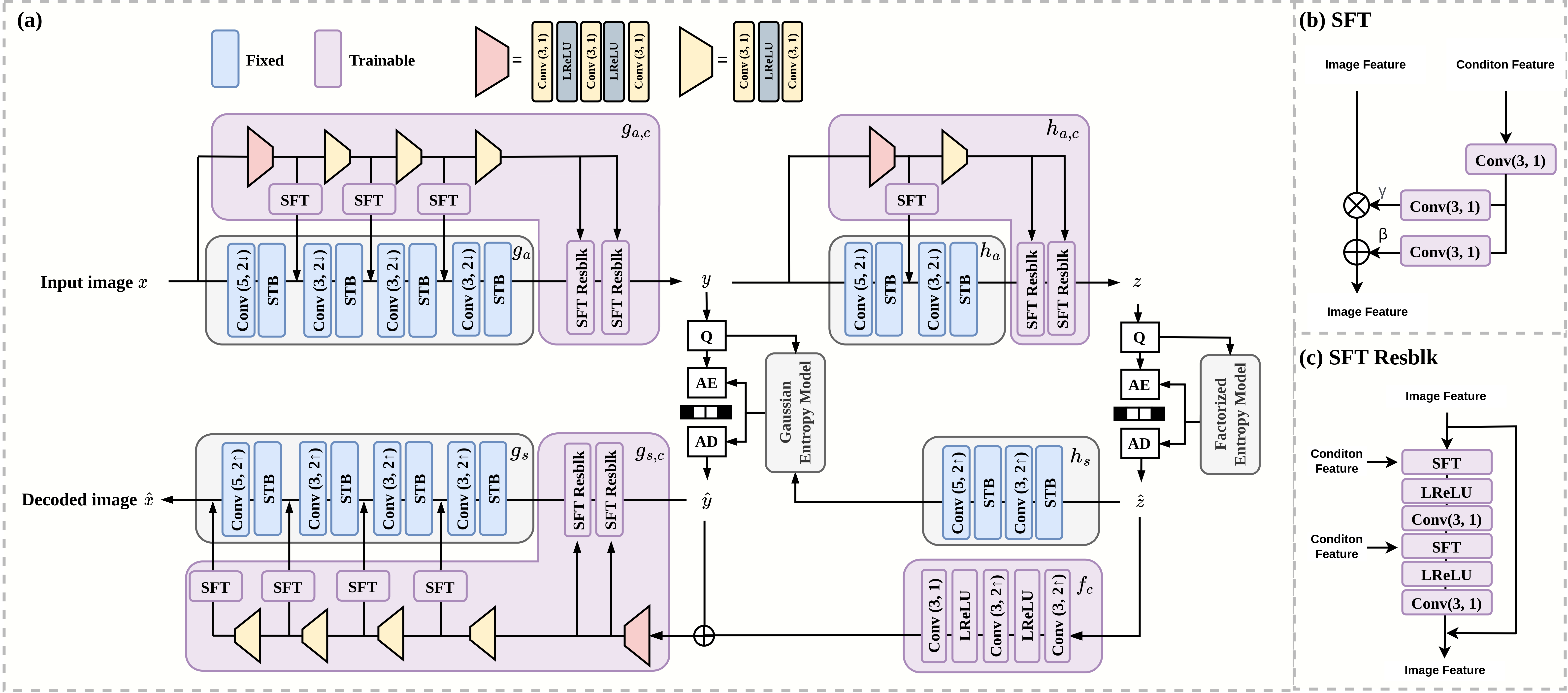}
  \caption{Architecture of \textit{TIC+SFT}.
  }
  \label{fig:ICCV}
\end{figure*}
\begin{figure}[t!]
  \centering
  \includegraphics[width=0.99\linewidth]{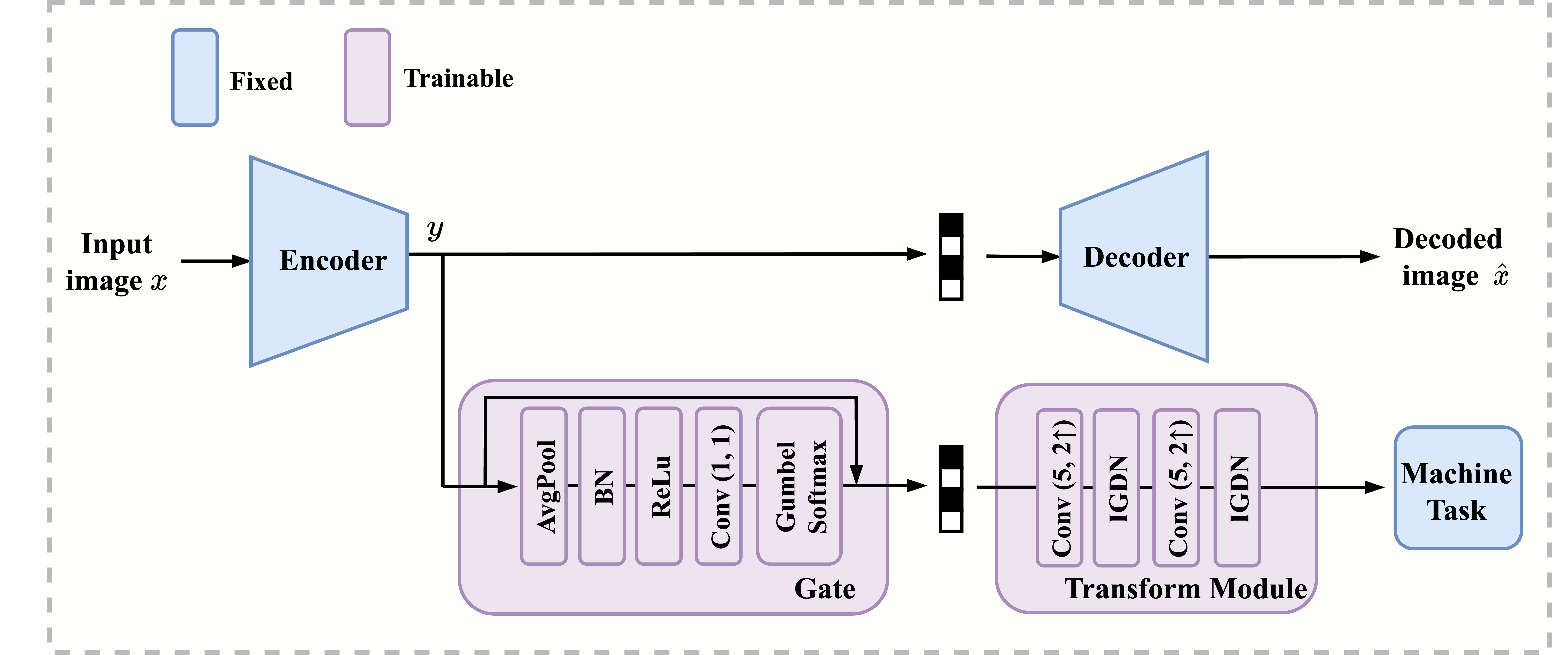}
  \caption{Architecture of \textit{TIC+channel selection}.}
  \label{fig:ICPR}
\end{figure}

\subsection{Extractor in Prompt Generator}
Fig.~\ref{fig:unet} details the network architecture of the extractor in our task-specific prompt generator $g_p$ (see Fig.~2(a) in the main paper). It has a U-Net~\cite{olaf2015unet}-like structure. 

\subsection{\textbf{\textit{TIC+SFT}}}
Fig.~\ref{fig:ICCV} depicts the network architecture of the baseline method \textit{TIC+SFT}~\cite{wang2018sftgan}, which shares the same fixed pre-trained base codec (the parts in blue color) as our \textit{TransTIC}. \textit{TIC+SFT} utilizes spatial feature transform (SFT) layers to perform element-wise affine transformation of the feature maps in $g_a, g_s$, and $h_a$ for transferring the base codec from human perception to downstream machine tasks. It follows~\cite{ICCV21} in using convolutional neural networks to produce the element-wise affine parameters $\gamma, \beta$ for each SFT layer.

\subsection{\textbf{\textit{TIC+channel selection}}}
Fig.~\ref{fig:ICPR} shows the architecture of \textit{TIC+channel selection}~\cite{channelselection}. Based on a pre-trained codec for human perception, \textit{TIC+channel selection} introduces two additional task-specific modules for machine perception. \textcolor{black}{As shown, a gate module first performs adaptive channel selection on the image latent $y$ through multiplying each of its channels by a binary value. Then, a transform module converts the masked image latent into a set of feature maps suitable for the downstream recognition network.}
\begin{figure*}[t]
\centering
\subfigure[Human Perception]{
\centering
\includegraphics[width=0.235\linewidth]{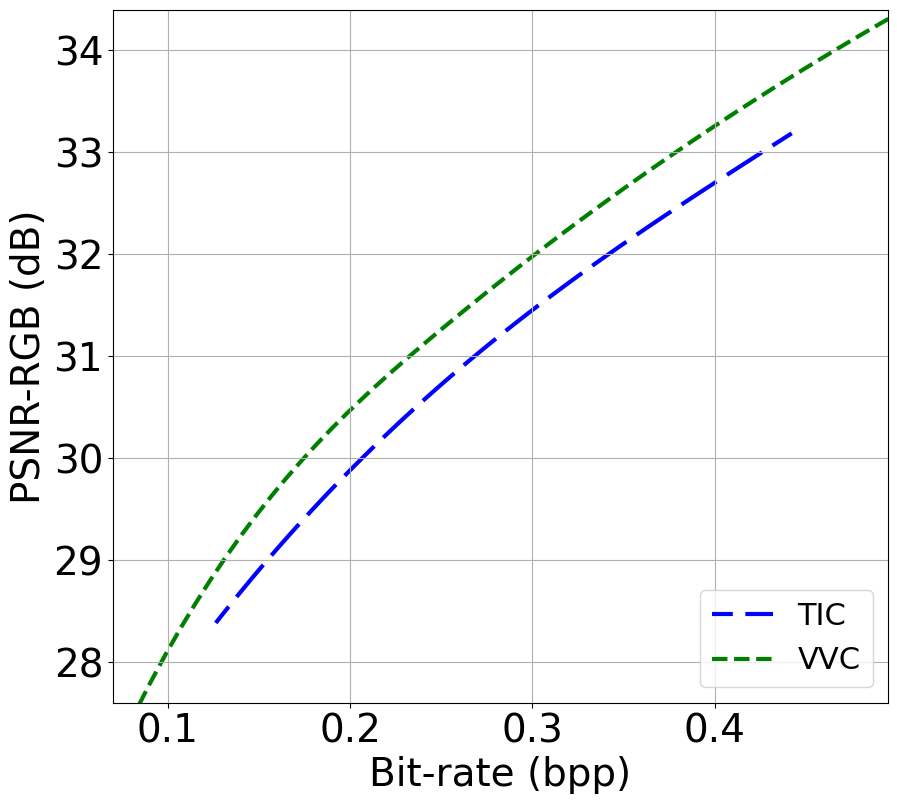}
}
\subfigure[Classification]{
\centering
\includegraphics[width=0.235\linewidth]{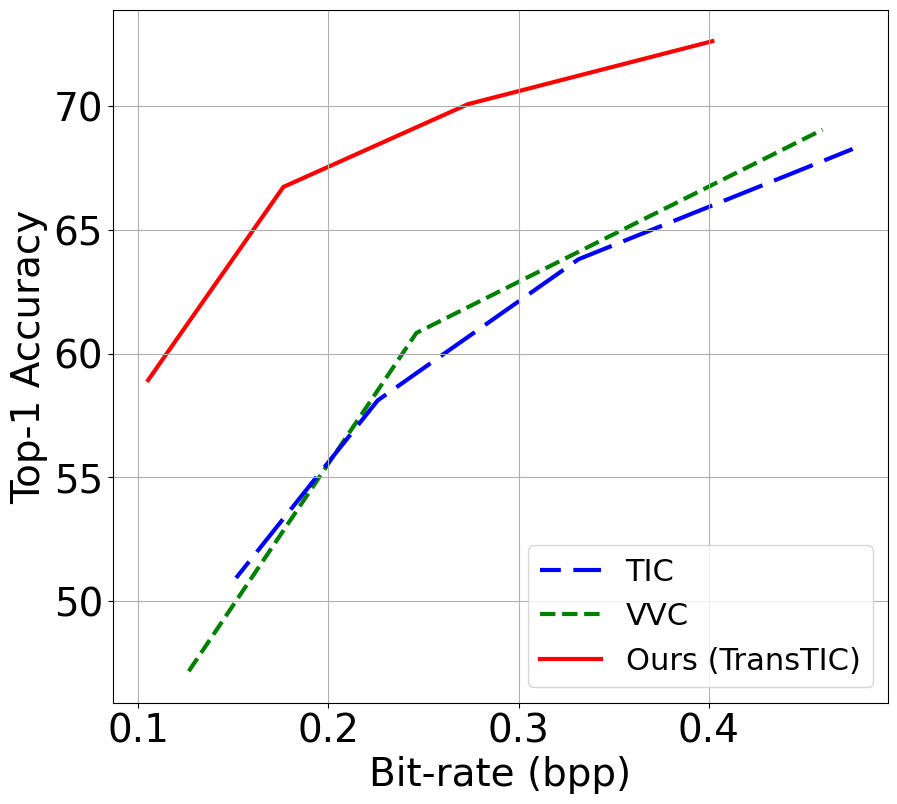}
}
\subfigure[Object Detection]{
\centering
\includegraphics[width=0.235\linewidth]{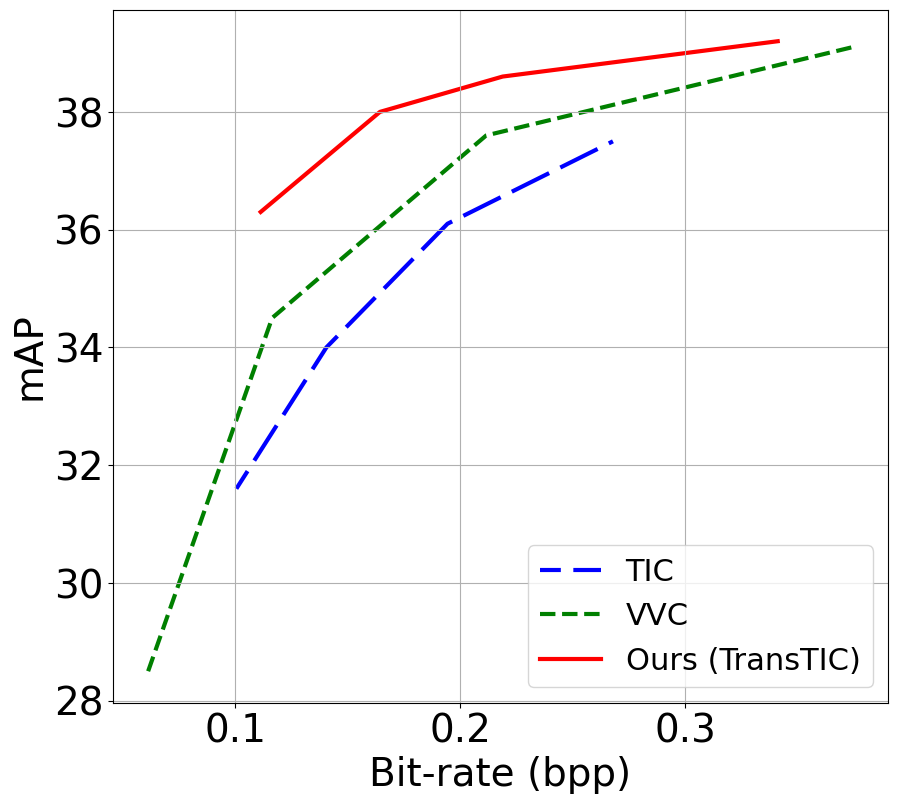}
}
\subfigure[Instance Segmentation]{
\centering
\includegraphics[width=0.235\linewidth]{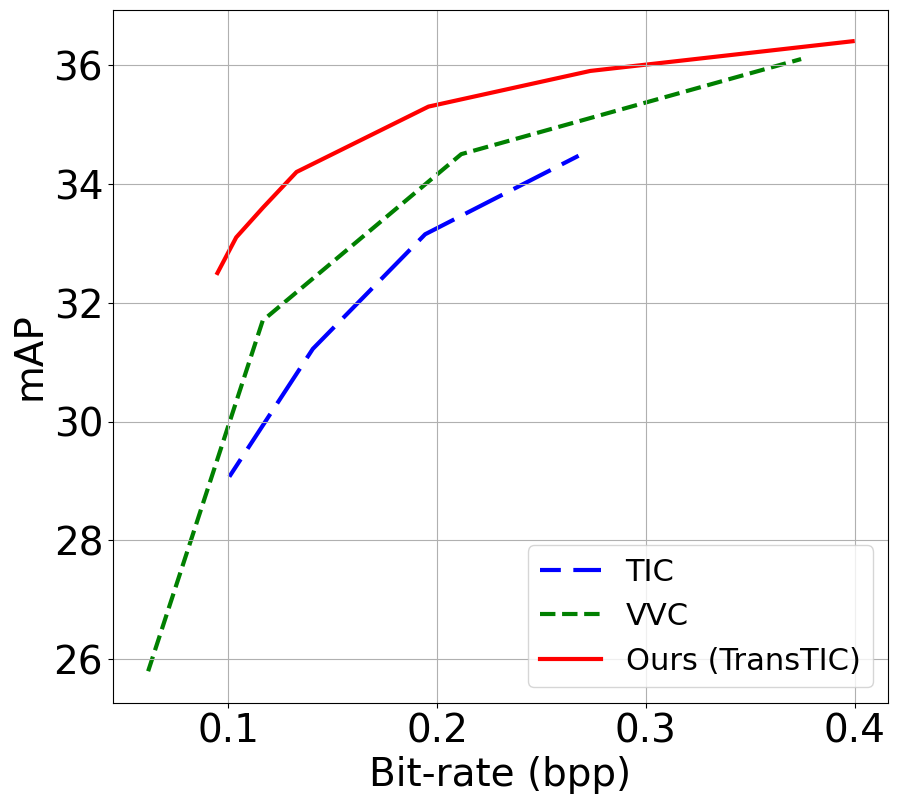}
}
\caption{Performance comparison between our \textit{TransTIC} and VVC under various tasks.}
\label{fig:RD_vvc}
\end{figure*}
\section{Comparison with VVC}
\label{sec:VVC}
Fig.~\ref{fig:RD_vvc} (a) compares our base codec, \textit{TIC}, with the state-of-the-art traditional codec VVC (VTM 16.0 intra coding) on the standard image compression task (i.e.~for human perception). The dataset is Kodak~\cite{Kodak}. As shown, \textit{TIC} shows worse PSNR results than VVC on the standard reconstruction task. It is thus not surprising to see that \textit{TIC} performs worse than VVC on the remaining recognition tasks. However, based on~\textit{TIC}, our~\textit{TransTIC} achieves much better rate-accuracy performance than VVC (Fig.~\ref{fig:RD_vvc} (b)(c)(d)). This result confirms the effectiveness of our prompting technique.    


\section{More Ablation Experiments}
\label{sec:ablation_supp}
\subsection{Prompt Injection: Deep vs. Shallow}
\label{sec:ablation_position}
\begin{figure}[t]
  \centering
  \includegraphics[width=0.85\linewidth]{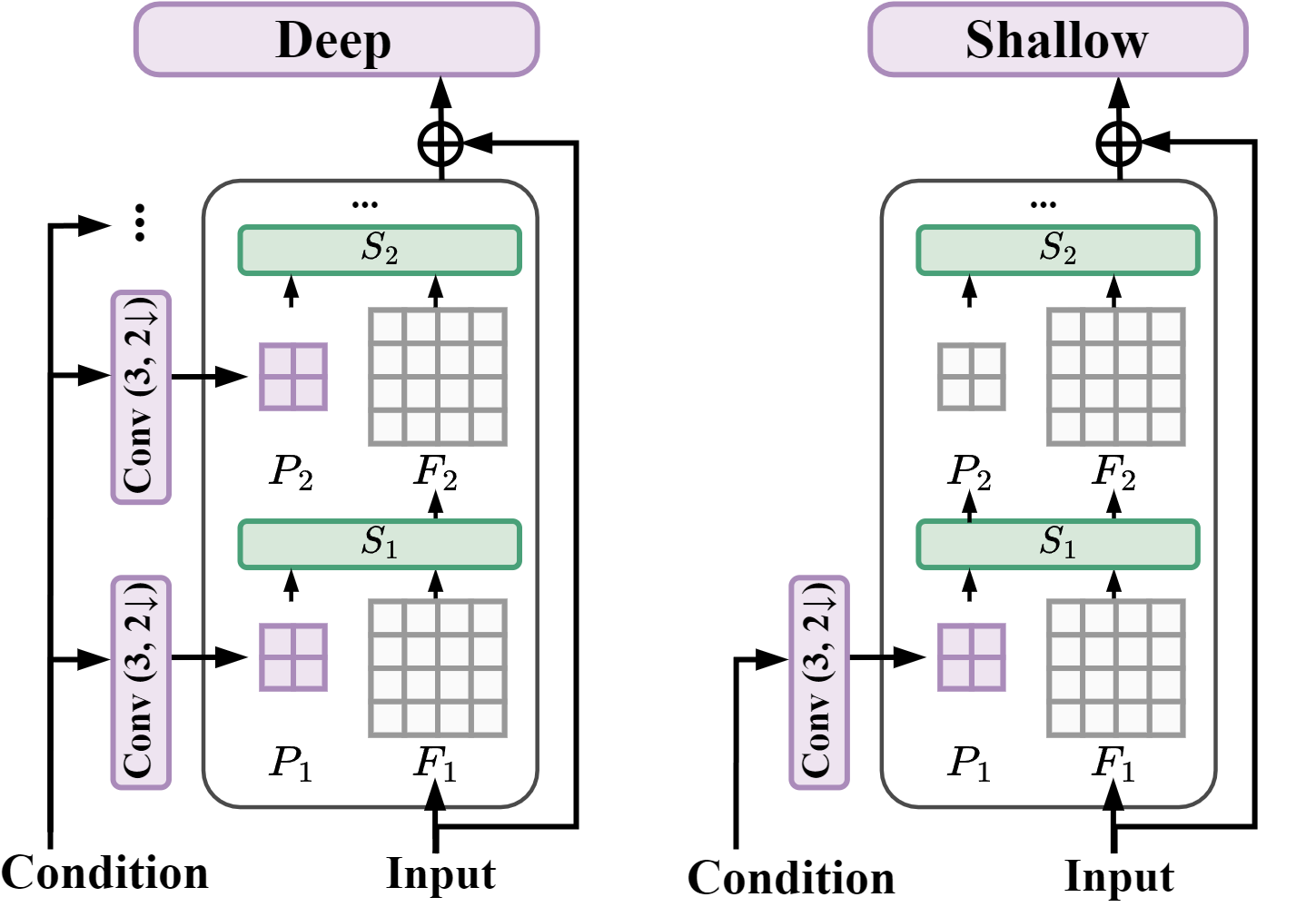}
  \caption{Architecture comparison of \textit{Deep} and \textit{Shallow} IP-type STB.}
  \label{fig:deep_vs_shallow}
  \vspace{-0.2cm}
\end{figure}

\begin{figure}[t]
\centering
\subfigure[Classification]{
\centering
\includegraphics[width=0.48\linewidth]{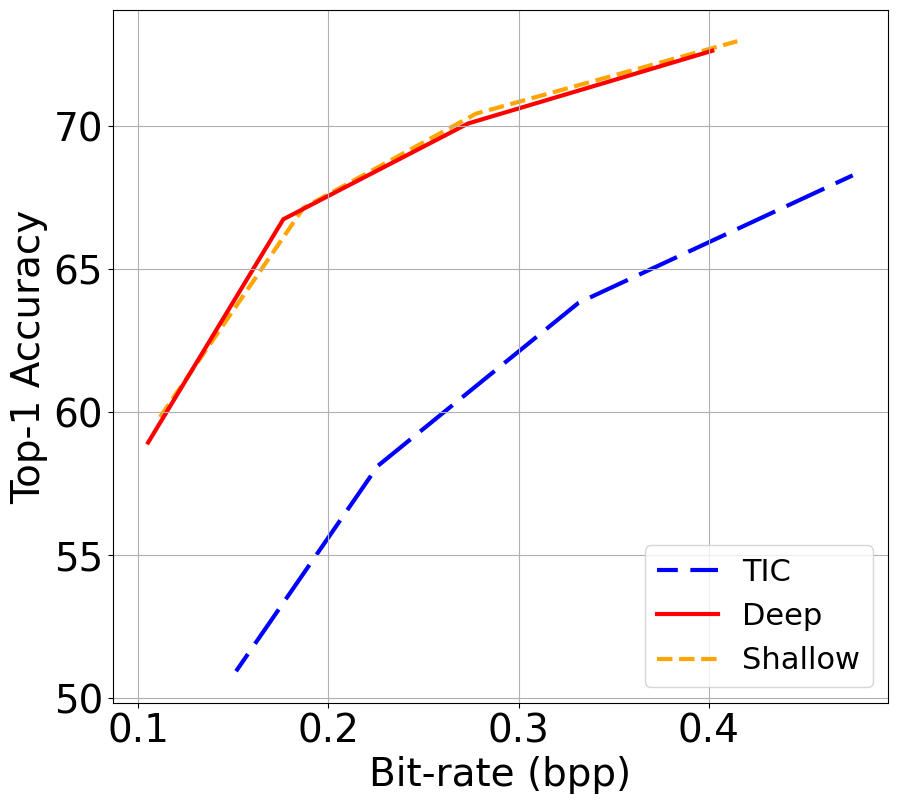}
}
\hspace{-0.35cm}
\subfigure[Object Detection]{
\centering
\includegraphics[width=0.49\linewidth]{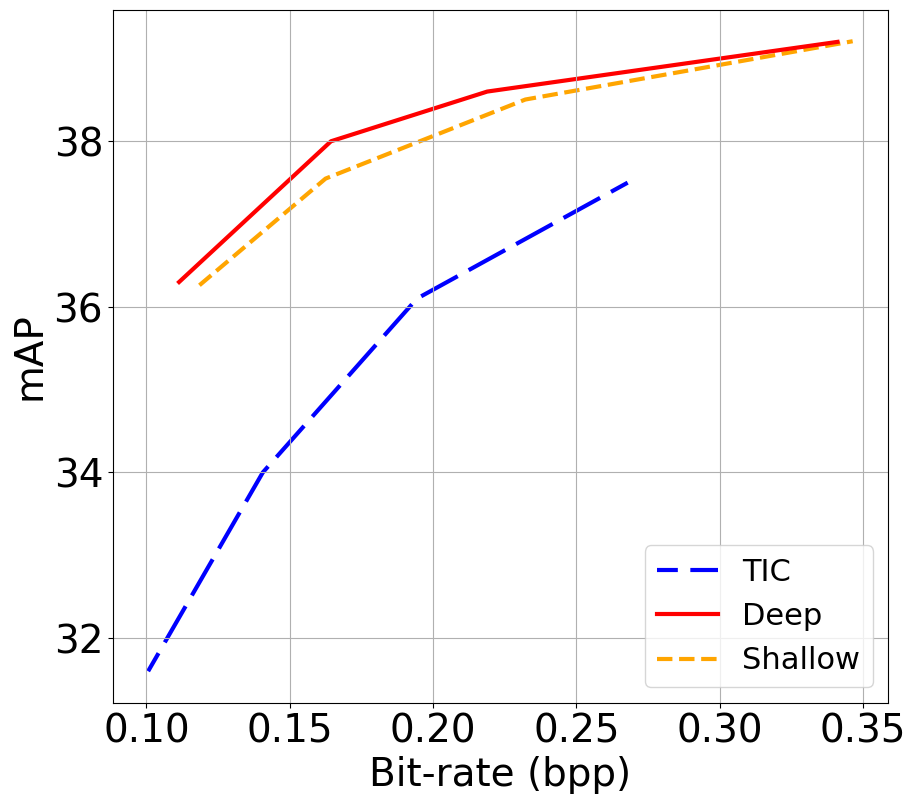}
}
\caption{Ablation on prompt injection: \textit{Deep} vs. \textit{Shallow}.}
\label{fig:ablation_injection_position}
\end{figure}

This ablation experiment tests another variant of prompt injection. Our~\textit{TransTIC} injects prompts to every Swin-Transformer layer in an IP-type or TP-type STB, which is similar to VPT-Deep in~\cite{VPT}. Another possible way of injecting prompts is to insert them only at the first Swin-Transformer layer of a STB. These prompts are also updated in the multi-head self-attention step. This setting is analogous to VPT-shallow in \cite{VPT}. The architectural difference between \textit{Deep} and \textit{Shallow} is shown in Fig.~\ref{fig:deep_vs_shallow}. From Fig.~\ref{fig:ablation_injection_position}, \textit{Deep} performs comparably to \textit{Shallow} on the classification task, and performs slightly better than \textit{Shallow} on the detection task. In Table~\ref{tab:complexity_ablation}, \textit{Deep} has comparable kMAC/pixel and model size to \textit{Shallow}. We thus choose \textit{Deep} in our \textit{TransTIC} for its better rate-accuracy performance.



\subsection{IP-type STBs in the Decoder}
\label{sec:ablation_dec_IP}
\begin{figure}[t]
  \centering
  \includegraphics[width=0.9\linewidth]{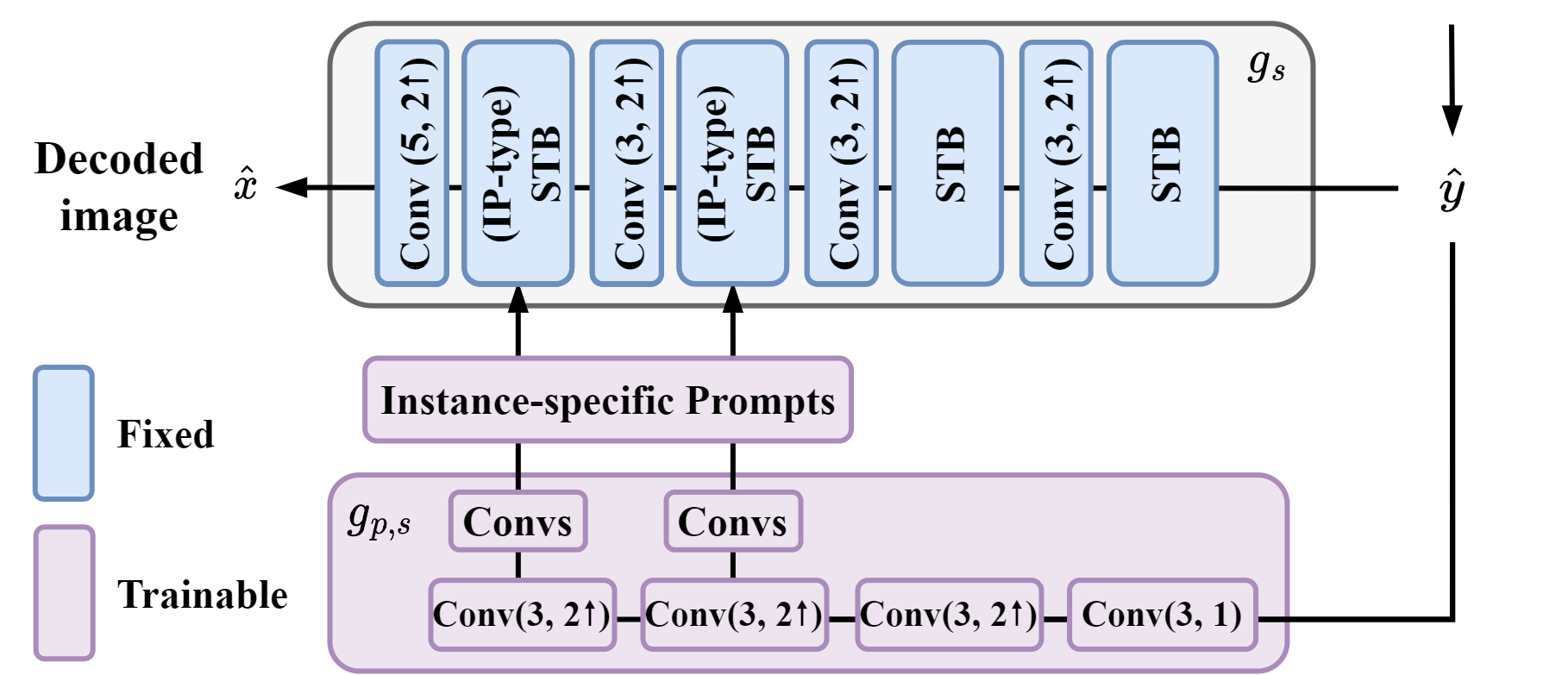}
  \caption{Architecture of IP-type STBs in the decoder.}
  \label{fig:dec_IPSTB}
\end{figure}

\begin{figure}[t]
\centering
\subfigure[Classification]{
\centering
\includegraphics[width=0.48\linewidth]{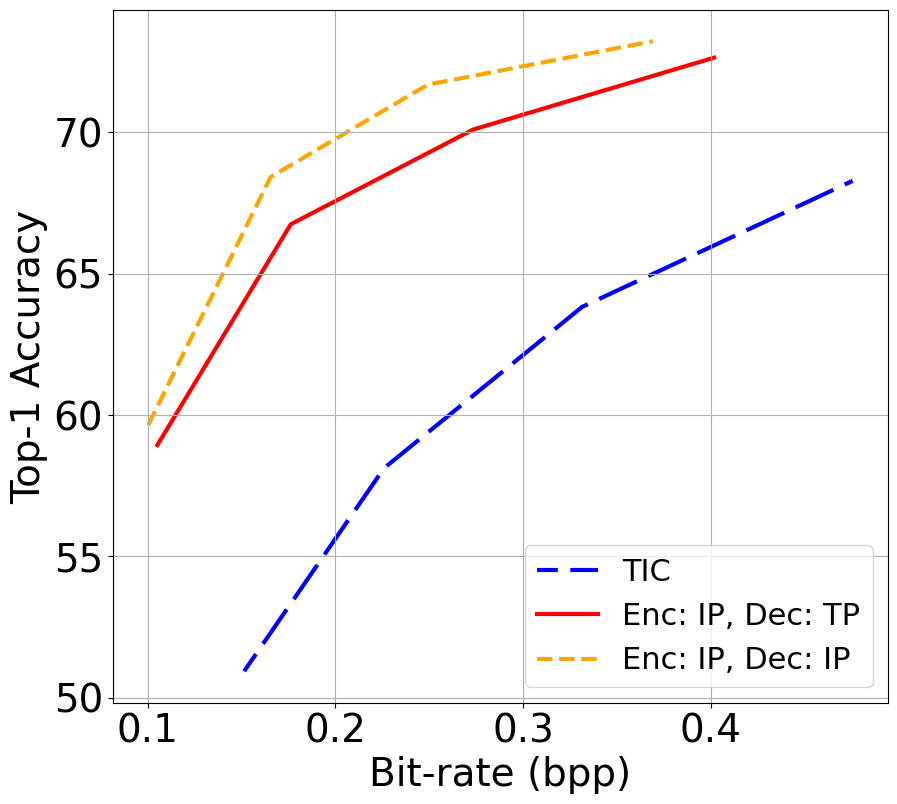}
}
\hspace{-0.35cm}
\subfigure[Object Detection]{
\centering
\includegraphics[width=0.48\linewidth]{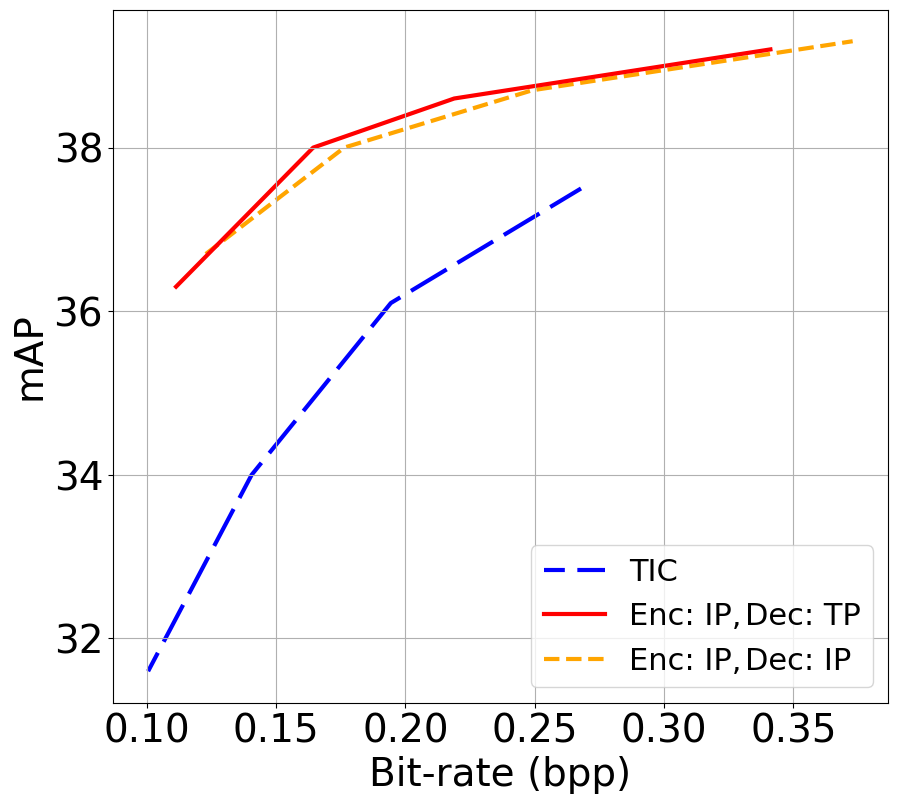}
}
\caption{Ablation on decoder side STB.}
\label{fig:ablation_IP_dec}
\vspace{-0.2cm}
\end{figure}

This ablation study introduces IP-type STBs to the decoder. Currently, our \textit{TransTIC} uses only TP-type STBs in the decoder because the input image is not accessible on the decoder side. One alternative to constructing IP-type STBs on the decoder side is to utilize the decoded latent $\hat{y}$ to generate instance-specific prompts (Fig.~\ref{fig:dec_IPSTB})). From Fig.~\ref{fig:ablation_IP_dec}, we see that introducing such IP-type STBs to the decoder improves the rate-accuracy performance on the classification task, but performs comparably to TP-type STBs on the object detection task. From Table~\ref{tab:complexity_ablation}, as compared to TP-type STBs, IP-type STBs lead to a 36\% increase in the decoder's kMACs/pixel and a 30\% increase in the decoder's model size. Because low decoding complexity and small decoder size are of importance, we choose to use TP-type STBs in the decoder.



\subsection{Prompt Numbers} 
\label{sec:ablation_number}
\begin{figure}[t]
\centering
\subfigure[Classification]{
\centering
\includegraphics[width=0.48\linewidth]{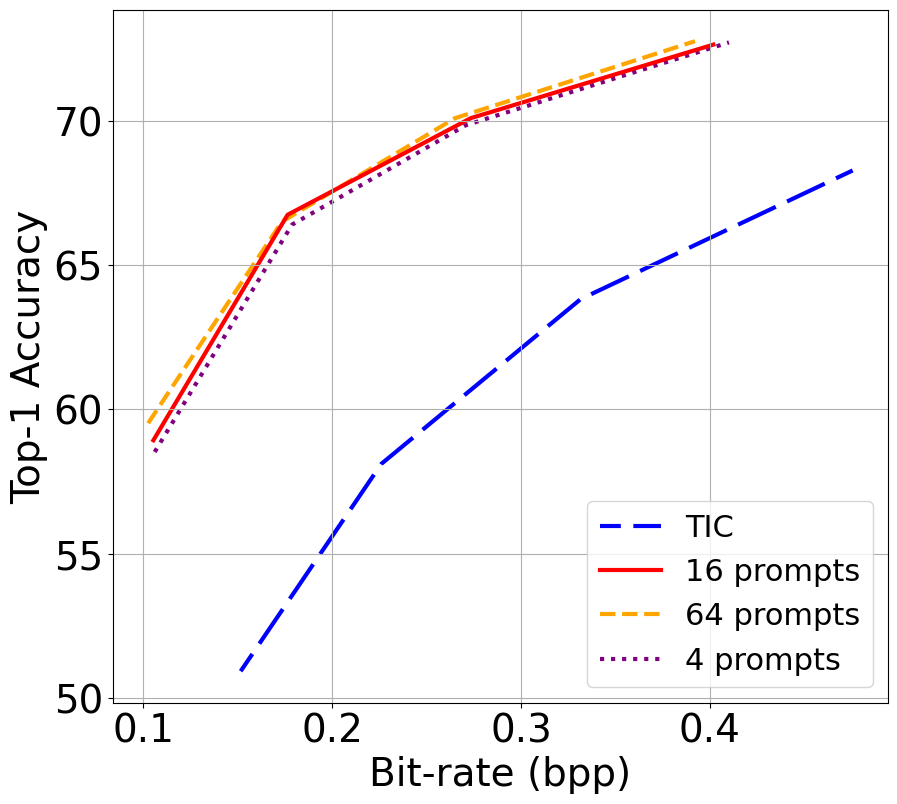}
}
\hspace{-0.35cm}
\subfigure[Object Detection]{
\centering
\includegraphics[width=0.49\linewidth]{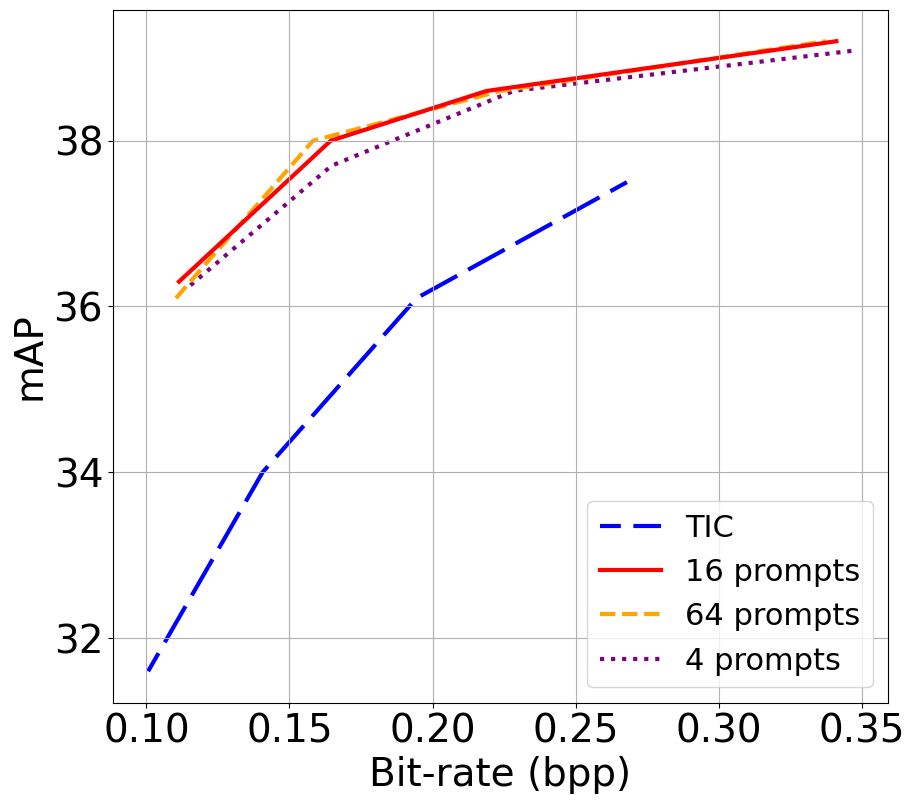}
}
\caption{Ablation on the number of prompts.}
\label{fig:ablation_number}
\end{figure}

Fig.~\ref{fig:ablation_number} ablates the effect of the number of prompts used in a Swin-Transformer window. When the number of prompts decreases from 64 to 4, the rate-accuracy performance drops marginally on the more complicated detection task. According to Table~\ref{tab:complexity_ablation}, the kMACs/pixel and model size of the model with 16 prompts is close to those of the model with 4 prompts. We thus choose 16 prompts to strike a balance between the rate-accuracy performance and model complexity. 

\begin{table}[]
\caption{Comparison of the kMACs/pixel and model size. \textbf{Bold} indicates our final design choices.}
\setlength{\tabcolsep}{4.0pt}
\fontsize{8}{8}\selectfont
\centering
\begin{tabular}{c|c|cc|cc}
\toprule
                         \multirow{2}{*}{Section} & \multirow{2}{*}{Method} & \multicolumn{2}{c}{\textbf{kMACs/pixel}} & \multicolumn{2}{|c}{\textbf{Params (M)}} \\ 
                              &  & Encoder             & Decoder            & Encoder            & Decoder            \\ 
\midrule
& \textit{TIC}                                              & 142.54              & 188.52             & 3.65               & 3.86               \\
\midrule
\multirow{2}{*}{\ref{sec:ablation_position}}  
& \textit{Shallow}                        & 322.80              & 209.51             & 4.65               & 3.88               \\
& \textit{\textbf{Deep}}                  & 332.03              & 202.60             & 5.24               & 3.89               \\
\midrule
\multirow{2}{*}{\ref{sec:ablation_dec_IP}}  
 & Enc: IP, Dec: IP               & 332.03              & 276.39             & 5.24               & 5.06               \\
 & \textbf{Enc: IP, Dec: TP}               & 332.03              & 202.60             & 5.24               & 3.89               \\
\midrule
\multirow{3}{*}{\ref{sec:ablation_number}} 
& 4 prompts                                                 & 302.06              & 192.04             & 5.24               & 3.87               \\
& \textbf{16 prompts}                                       & 332.03              & 202.60             & 5.24               & 3.89               \\
& 64 prompts                                                & 451.91              & 244.87             & 5.24               & 3.98               \\
\midrule
\multirow{3}{*}{\ref{sec:ablation_depth}} 
& \textbf{STB-1234}                            & 332.03              & 202.60             & 5.24               & 3.89               \\
& STB-12                                       & 332.03              & 200.80             & 5.24               & 3.87               \\
& STB-34                                       & 332.03              & 190.32             & 5.24               & 3.88               \\
\midrule
\multirow{3}{*}{\ref{sec:ablation_enc_vs_dec}} 
 & Enc: IP, Dec: -               & 332.03              & 188.52             & 5.24               & 3.86               \\
 & Enc: - , Dec: TP              & 142.54              & 202.60             & 3.65               & 3.89               \\
 & \textbf{Enc: IP, Dec: TP}     & 332.03              & 202.60             & 5.24               & 3.89               \\
\bottomrule
\end{tabular}
\label{tab:complexity_ablation}
\end{table}

\subsection{Prompt Depth of the Decoder}
\label{sec:ablation_depth}
\begin{figure}[t]
\centering
\subfigure[Classification]{
\centering
\includegraphics[width=0.47\linewidth]{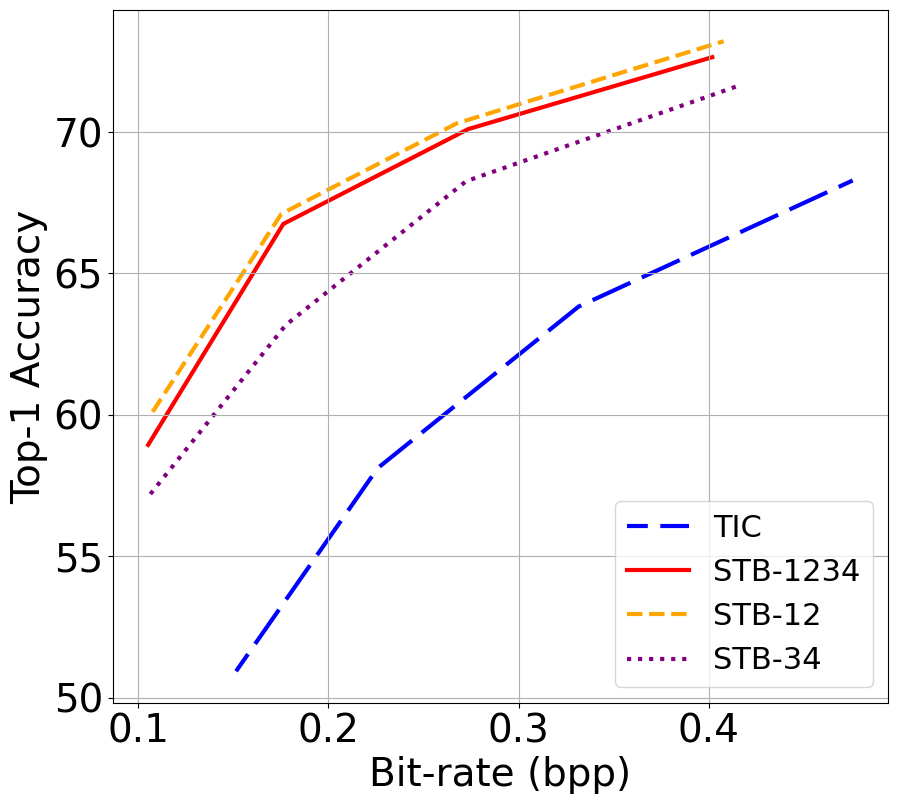}
}
\hspace{-0.35cm}
\subfigure[Object Detection]{
\centering
\includegraphics[width=0.49\linewidth]{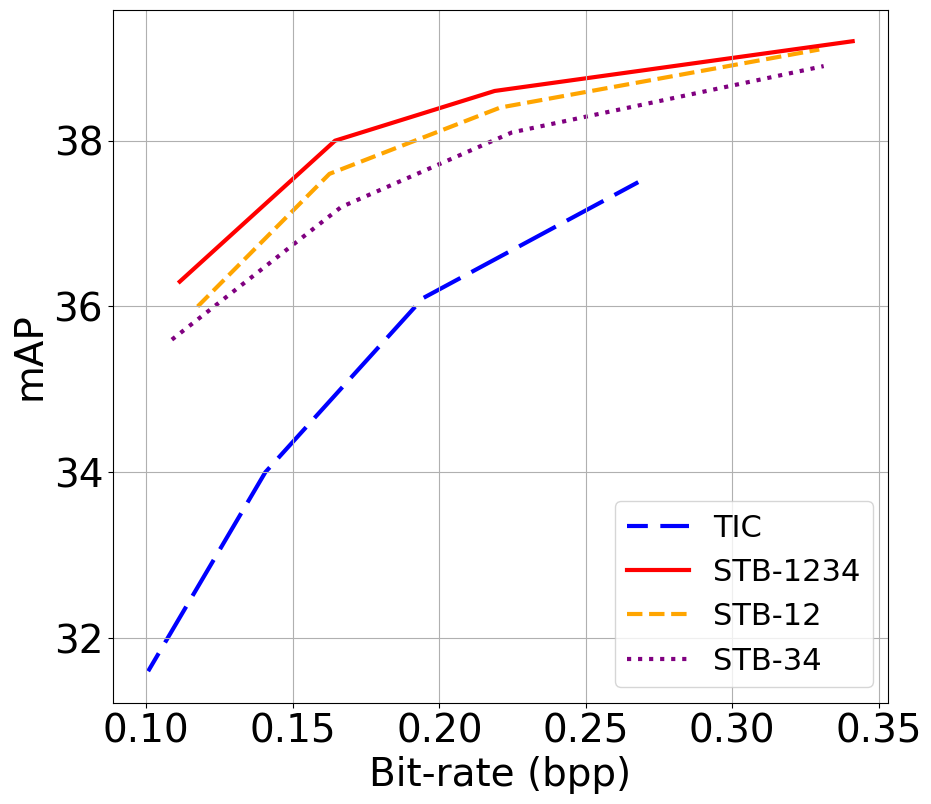}
}
\caption{Ablation on the prompt depth of the decoder.}
\label{fig:ablation_depth}
\end{figure}

Fig.~\ref{fig:ablation_depth} analyzes which and how many STBs to inject prompts on the decoder side. As shown, injecting task-specific prompts to all STBs (STB-1234) appears to be a better choice than the other variants, namely, STB-12 and STB-34, in terms of the rate-accuracy performance. STB-12 refers to injecting prompts to the two STBs closer to the decoded image~$\hat{x}$ while STB-34 refers to injecting them to STBs closer to the image latent. From Fig.~\ref{fig:ablation_depth}, STB-12 performs better than STB-34. Because STB-1234 has only slightly higher kMAC/pixel and model size than STB-12 (Table~\ref{tab:complexity_ablation}), we choose STB-1234 as our final design. 

\subsection{Prompting Encoder vs. Decoder}
\label{sec:ablation_enc_vs_dec}
\begin{figure}[t]
\centering
\subfigure[Classification]{
\centering
\includegraphics[width=0.47\linewidth]{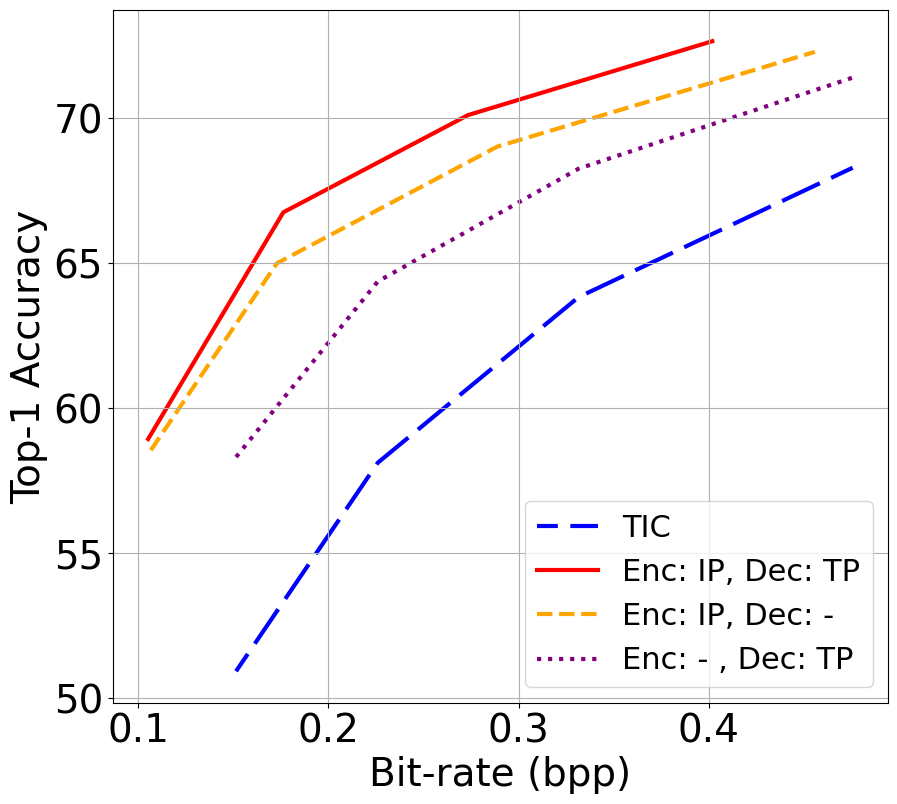}
}
\hspace{-0.35cm}
\subfigure[Object Detection]{
\centering
\includegraphics[width=0.49\linewidth]{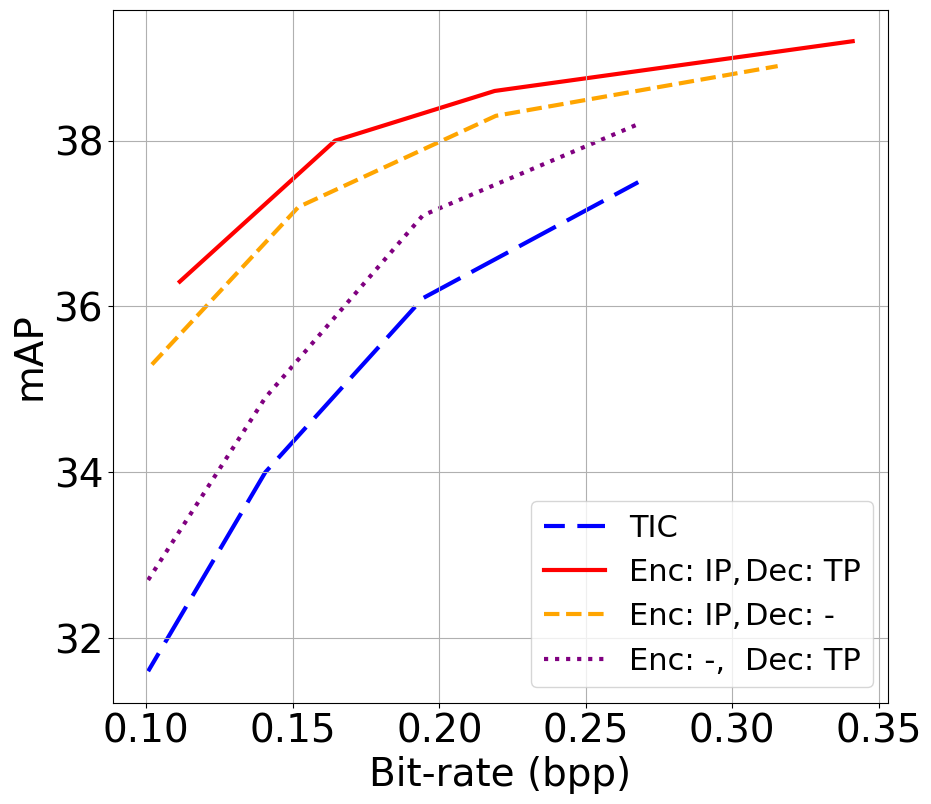}
}
\caption{Ablation on effectiveness of prompt on encoder and decoder sides.}
\label{fig:enc_vs_dec}
\end{figure}

Fig.~\ref{fig:enc_vs_dec} compares the effectiveness of introducing IP-type STBs to the encoder and TP-type STBs to the decoder. As shown, introducing prompts to both the encoder and decoder achieves the best rate-accuracy performance. We also see that prompting on the encoder side is more effective than prompting on the decoder side. This result is intuitively agreeable because prompting on the encoder side allows the compressed bitstream to be tailored for the downstream task. The complexity characteristics of these variants are provided in Table~\ref{tab:complexity_ablation}.

\section{More Qualitative Results}
\label{sec:qualitative_supp}
\begin{figure*}[t]
\centering
\subfigure[]{
\centering\
\includegraphics[width=0.99\linewidth]{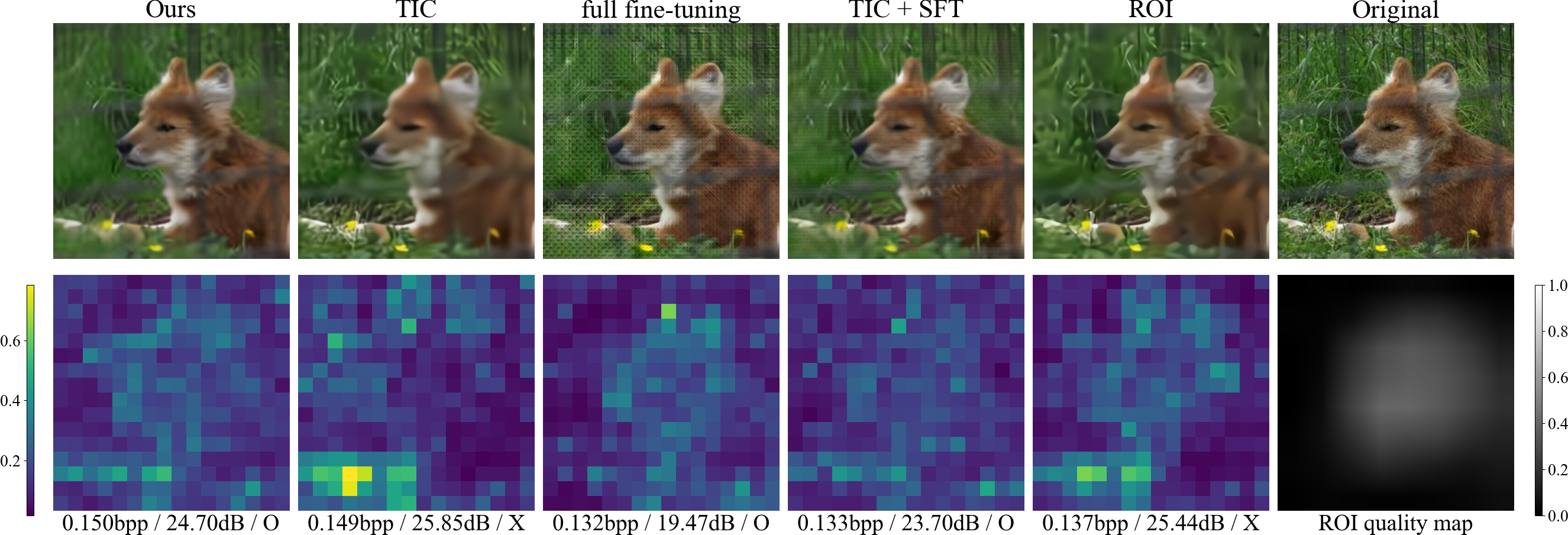}
}
\subfigure[]{
\centering\
\includegraphics[width=0.99\linewidth]{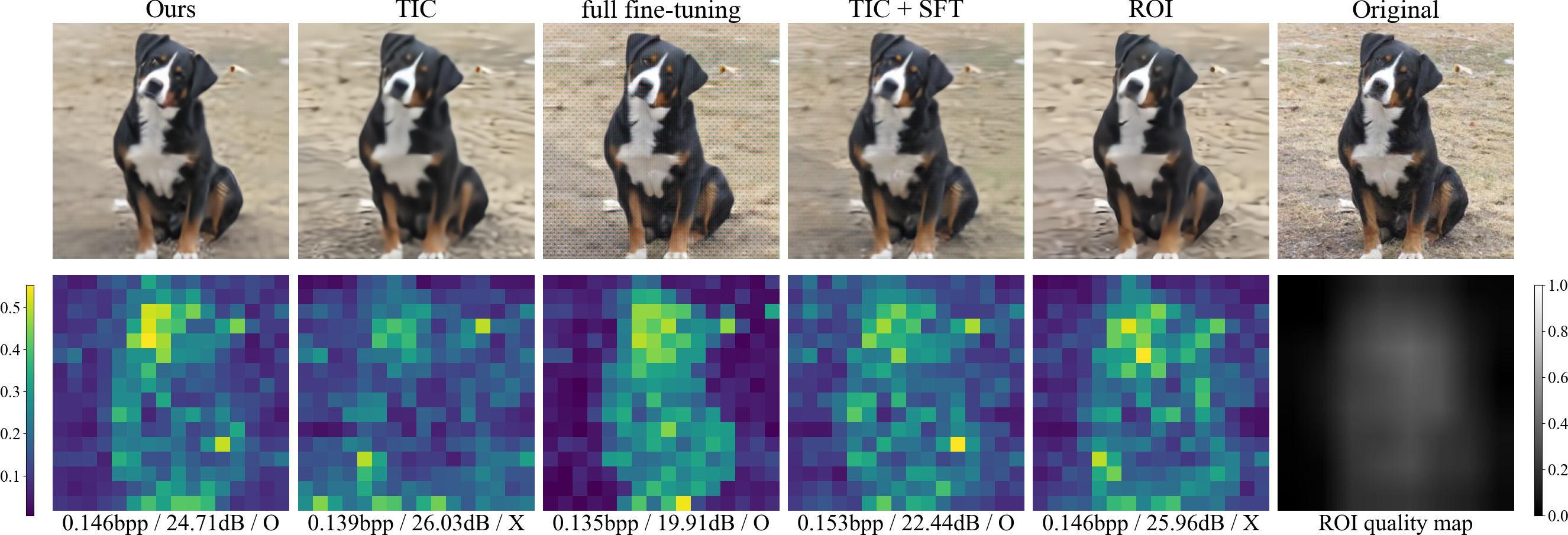}
}
\caption{Visualization of the decoded images (the first row) and the bit allocation maps (the second row) of the image latent $\hat{y}$ for the classification task. The rightmost image of the second row shows the quality map used for the ROI method. The text below each map denotes the corresponding bit rate / PSNR / prediction result, with O and X indicating correct and false classification, respectively.}
\label{fig:visualization_cls}
\vspace{-1mm}
\end{figure*}

\begin{figure*}[t]
\centering
\subfigure[]{
\centering\
\includegraphics[width=0.99\linewidth]{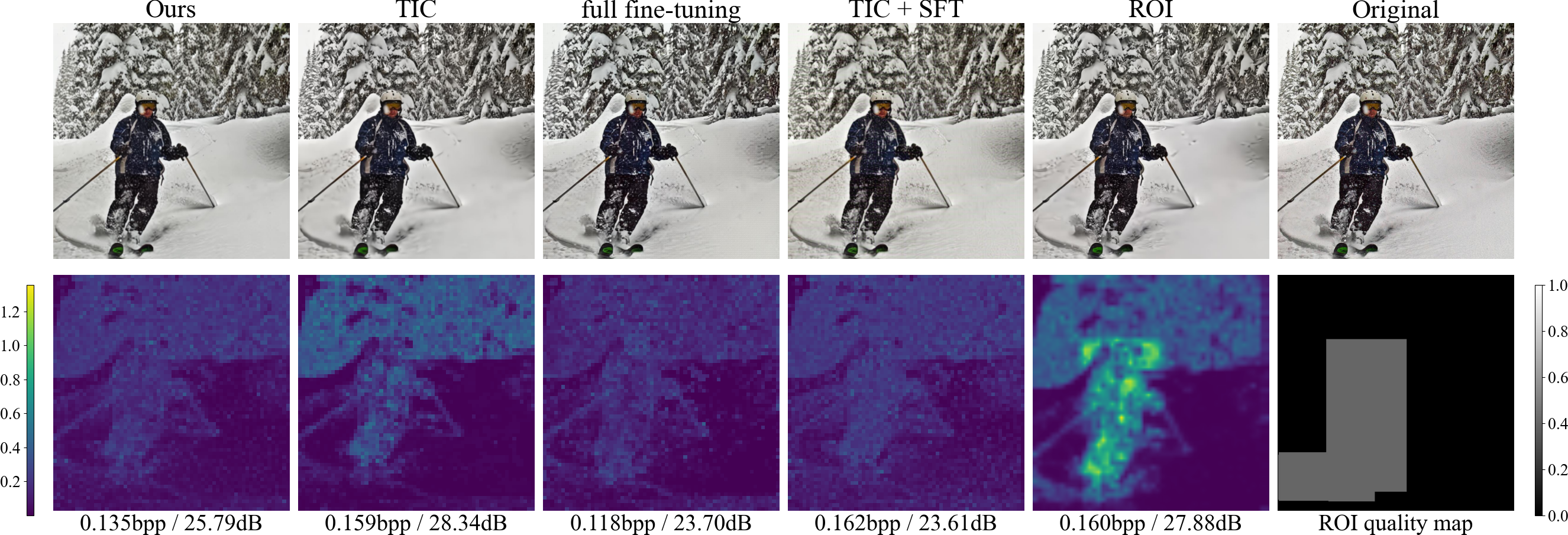}
}
\subfigure[]{
\centering\
\includegraphics[width=0.99\linewidth]{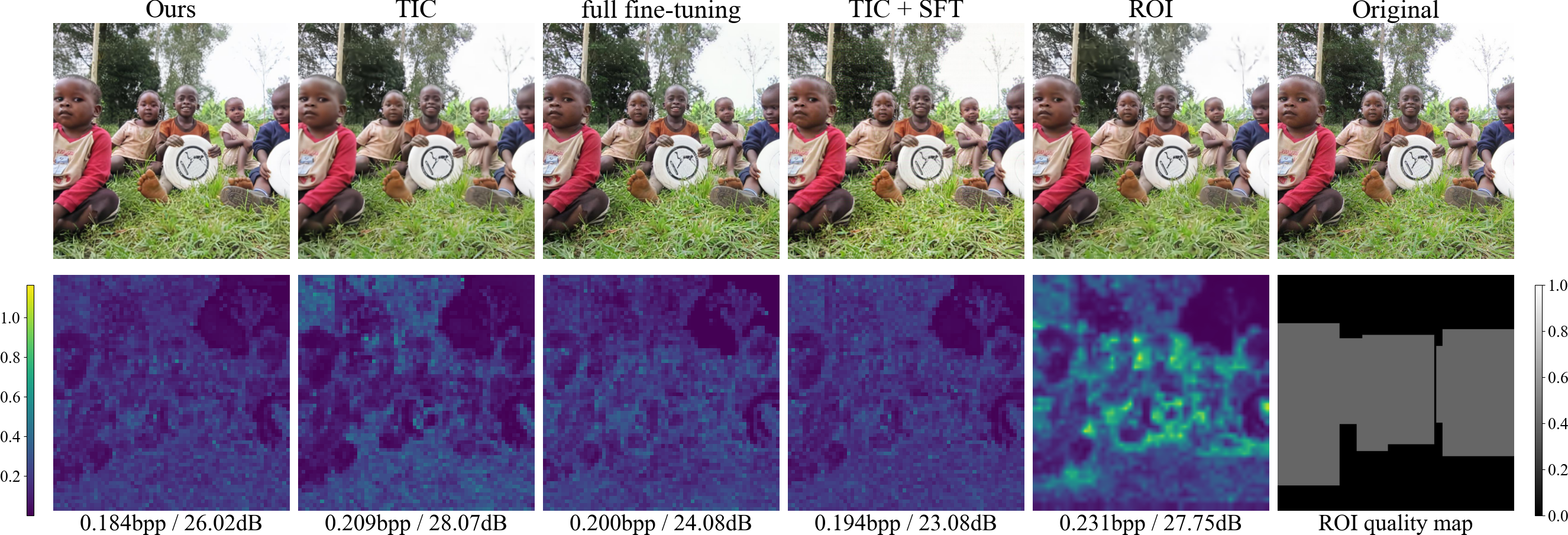}
}
\caption{Visualization of the decoded images (the first row) and the bit allocation maps (the second row) of the image latent $\hat{y}$ for the object detection task. The rightmost image of the second row shows the quality map used for the ROI method. The text below each map denotes the corresponding bit rate / PSNR.}
\label{fig:visualization_det}
\vspace{-1mm}
\end{figure*}

\begin{figure*}[t]
\centering
\subfigure[]{
\centering\
\includegraphics[width=0.99\linewidth]{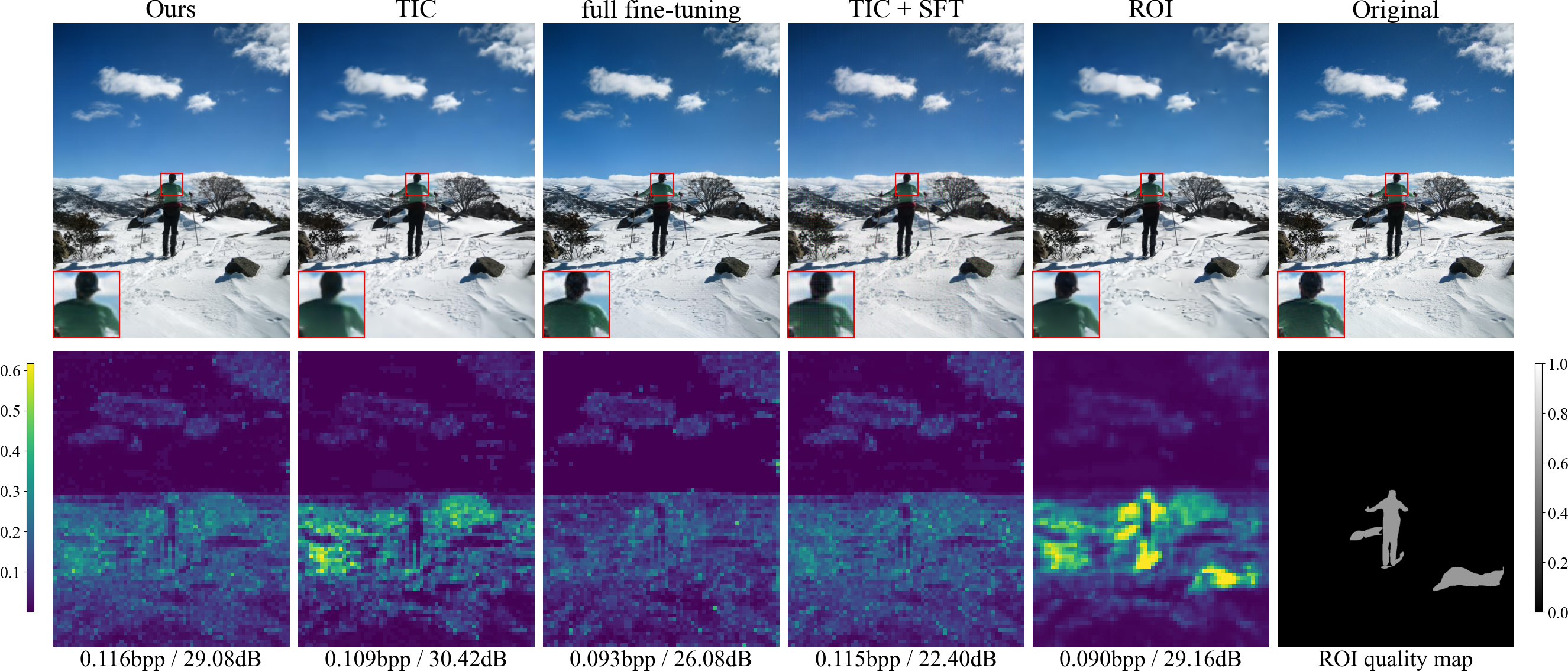}
}
\subfigure[]{
\centering\
\includegraphics[width=0.99\linewidth]{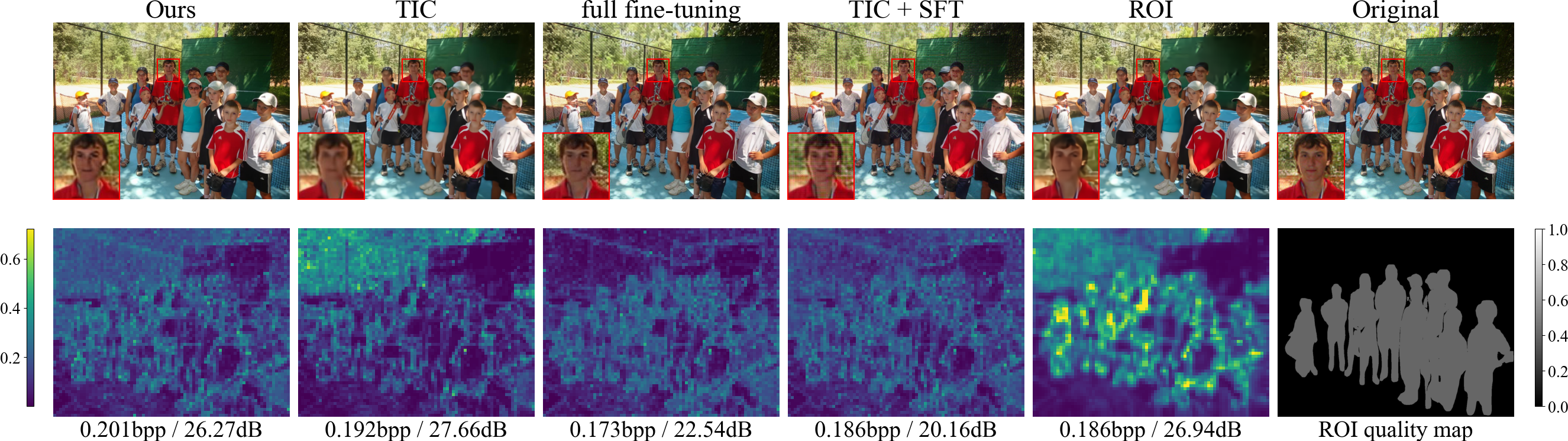}
}
\caption{Visualization of the decoded images (the first row) and the bit allocation maps (the second row) of the image latent $\hat{y}$ for the instance segmentation task. The rightmost image of the second row shows the quality map used for the ROI method. The text below each map denotes the corresponding bit rate / PSNR.}

\label{fig:visualization_seg}
\vspace{-1mm}
\end{figure*}

Fig.~\ref{fig:visualization_cls}, Fig.~\ref{fig:visualization_det}, and Fig.~\ref{fig:visualization_seg} provide more qualitative results, comparing the decoded images and the bit allocation maps produced by the competing methods. As shown, \textit{TIC}, the codec optimized for human perception, tends to allocate more bits to complex regions, even if those regions are less relevant (e.g. background) to the downstream recognition tasks. In contrast, the other methods, which target machine perception, attempt to shift coding bits from the background regions to the foreground objects.



\clearpage

\end{document}